\documentclass[a4paper,12pt,oneside,sfdefaults=false]{scrartcl} 
\usepackage{float} 

\usepackage[backend=biber,style=apa,autocite=inline]{biblatex}
\DeclareLanguageMapping{english}{english-apa}
\usepackage[english]{babel}
\usepackage[utf8]{inputenc}
\usepackage{tgheros}

\usepackage[left=2.5cm,right=2.5cm,top=2.5cm,bottom=2.5cm]{geometry}

\usepackage[justification=RaggedRight, font=small, labelfont=bf, textfont=sl, format=hang, margin={1.5cm}]{caption}
\usepackage{amsmath} 

\usepackage{newpxtext}
\usepackage{newpxmath}

\usepackage{marvosym}

\usepackage{graphicx}
\usepackage{booktabs}
\usepackage{tabularx}
\usepackage{array}
\usepackage{multirow}
\usepackage{colortbl}
\usepackage{wrapfig}
\newcolumntype{L}[1]{>{\raggedright\arraybackslash}p{#1}}
\newcolumntype{C}[1]{>{\centering\arraybackslash}p{#1}}
\newcolumntype{R}[1]{>{\raggedleft\arraybackslash}p{#1}}

\usepackage{xcolor}
\definecolor{graylight}{rgb}{0.95,0.95,0.95}
\definecolor{green2}{RGB}{0, 167, 76}
\definecolor{blue3}{HTML}{5B51EB}
\definecolor{red1}{HTML}{F30000}

\usepackage{hyperref}
\hypersetup{
    colorlinks=true,
    linkcolor=blue,
    citecolor=blue,
    urlcolor=blue,
    pdftitle={Document Title},
    pdfauthor={Your Name},
    pdfsubject={Subject},
    pdfkeywords={Keyword1, Keyword2},
    pdfproducer={LaTeX with hyperref},
    pdfcreator={XeLaTeX, Overleaf}
}
\usepackage{bookmark}

\usepackage{lineno}
\modulolinenumbers[5]

\usepackage{fancybox}
\usepackage{setspace}
\usepackage{verbatim}
\usepackage{lipsum}
\usepackage{silence}
\usepackage[hang, flushmargin]{footmisc}
\graphicspath{{Figure/}}
\usepackage{csquotes}
\usepackage{rotating}

\AtEveryBibitem{\clearfield{isbn}\clearfield{issn}}
\AtEveryBibitem{%
    {
      \iffieldundef{year}
        {}
        {\clearfield{urldate}}
    }
    {}
}
\DeclareSourcemap{
  \maps[datatype=bibtex]{
    \map{
      \step[fieldsource=doi, final]
      \step[fieldset=url, null]
    }
  }
}

\title{
    \Large The Earth Moves, But So Does the Bias:
    Systematic Upward Bias of the Wasserstein (Earth Mover's)
    Distance and Permutation-Based Null Calibration
    \\[0.8em]
    {\large\textit{Accepted for publication in Political Analysis}}
}

\author{
    \large Ho Ting (Bosco) Hung\thanks{~~Department of Politics and International Relations, University of Oxford}
}

\date{\today}

\begin{document}

\maketitle

\setcounter{footnote}{0}

\normalsize
\begin{abstract}
    \textbf{Abstract:} The Earth Mover’s Distance (EMD) is gaining increasing interest among political scientists for assessing similarity in preference distributions. Existing applications predominantly involve relatively large samples and low-dimensional response spaces, which can mitigate finite-sample distortions. However, empirical EMD remains biased upward due to sampling variation in empirical probability measures. This problem is often overlooked and becomes especially severe in high-dimensional or sparse settings, including conjoint distributions that serve as an illustrative example in this paper. As political scientists are broadening their use cases of EMD, this paper cautions against interpreting standard bootstrap uncertainty bounds as a correction for the upward bias of empirical EMD. It proposes a permutation-based null calibration framework for more robust hypothesis testing. As a non-parametric approach, it frees researchers from making directional or distributional shape assumptions. While alternative estimators require these rigid assumptions to correct for upward bias, political science data often fail to meet them in practice. Through four sets of Monte Carlo simulations, this paper demonstrates the utility of this framework. The proposed approach also applies more generally to empirical comparisons of two probability distributions defined on a common metric space, provided that the ground distance between support points is substantively meaningful.
\end{abstract}

\newpage
\justifying
\doublespacing
\section{Introduction}
Evaluating congruence or preference distributions has been of key interest to political scientists \autocite{shim_measuring_2020}. The convergence and divergence of individual or group preferences underpin possibilities of conflicts and cooperation that drive political developments. To compare preference distributions, as \textcite{lupu_new_2017} survey, political scientists traditionally rely on computing the difference between the means, or comparing the overlap between the cumulative distribution functions of the preferences \autocite{golder_ideological_2010} or the probability distribution functions (Bhattacharyya coefficient) instead \autocite{bhattacharyya_measure_1946}. However, these methods can discard information about the shape and location of the distributions, require arbitrary quantization decisions, or be difficult to extend naturally beyond one dimension. 

A simple mass-elite congruence example illustrates the class of problems considered here. Suppose a researcher observes the ideological positions of 500 citizens and 50 elected representatives on a 0-10 left-right scale. Each group can be represented as an empirical probability distribution over 11 ordered positions. A natural question is how different these two distributions are: Do their means differ? How much probability mass would have to be shifted, and how far, to transform one distribution into the other? These questions also extend beyond this one-dimensional example. From ideology scaling to text-as-data and automated clustering applications, political scientists often summarize complex political phenomena as positions in multi-dimensional spaces \autocite{blumenau_citizens_2025, klar_multidimensional_2014, ramaciotti_american_2024, grimmer_general_2011}. More broadly, the two distributions being compared may be one- or multi-dimensional, continuous or discrete, may contain equal or unequal numbers of observations, or may come from responses to linear scale questions in simple surveys or alternative designs like factorial experiments. This brings the need for a flexible distance metric that is also capable of preserving meaningful information.

To address this, \textcite{lupu_new_2017} introduced the Earth Mover’s Distance (EMD/first Wasserstein Distance) to political science. Inspired by the problem of optimal transport \autocite{panaretos_statistical_2019, berger_wasserstein_2009}, it measures the difference between two probability distributions by calculating the minimum cost required to transform one into the other (shifting the mass or piles of `earth' to a matching set of holes) \autocite{kantorovich_mathematical_1960}. As it explicitly incorporates a ground distance between support points, it can measure both the amount of distributional difference and the distance over which that difference occurs. For example, suppose two distributions differ by the same amount of probability mass. Traditional bin-by-bin metrics (e.g., $L_1$ distance) and information-theoretic divergences (e.g., Jensen-Shannon divergence) do not account for the geometry of the support space. Therefore, unlike the EMD, they do not distinguish whether the mismatched mass lies at nearby or distant support points. However, when applied to political distributions, say ideology, this spatial awareness is important because shifting probability from one adjacent ideological position to another is less costly than shifting it across the full scale.

While receiving comparatively less attention from political scientists before the late 2010s \autocite{lupu_new_2017}, the EMD has since been increasingly applied to various substantive issues. Examples include cross-national differences in policy preferences \autocites{sorace_europeanisation_2026, fortunato_economic_2025}, and mass-elite/citizen-representative policy preferences \autocites{muriaas_attitudes_2025, broderstad_trustees_2025, carella_electoral_2024, sorace_does_2023, helliesen_unequal_2023, devine_convergence_2021, marzi_whom_2026, belschner_representation_2026}. These applications span different data structures, ranging from large one-dimensional survey samples to smaller subgroup comparisons and sparse distributions. Besides studying congruence, under the wave of large language models (LLMs) \autocite{bisbeeSyntheticReplacementsHuman2024, argyle_out_2023}, the EMD has also been applied by other scholars to evaluate how effectively synthetic agents can simulate human preferences concerning voting behaviour \autocite{ball_human_2025} and political beliefs or cultural values \autocite{cummins_threat_2026, cao_specializing_2025}.

Nonetheless, the EMD continues to suffer from an upward bias. Consider again the mass-elite example above: even if citizens and representatives were actually drawn from identical underlying ideological distributions, their two finite empirical distributions would rarely coincide exactly. Sampling variation alone can therefore generate a positive measured distance between them. This problem is common in naive plug-in estimators of non-negative distance metrics, where random noise accumulates rather than cancels out. It also exacerbates with the dimensionality of the metric space. In discrete settings, empirical distributions could become sparse and make empirical distributions appear distant even when drawn from identical parent populations.\footnote{Note that although this characterization is common, it does not apply to all distance metrics or estimators. For instance, the use of entropy estimation and smoothing can complicate the direction of bias.} 

While a few recent studies have acknowledged this issue and used bootstrapping to construct uncertainty bounds around their observed EMD estimates following the encouragement of \textcite{lupu_new_2017}, this can only capture the conditional sampling variability of the empirical statistic. Bootstrapping does not by itself identify the finite-sample null baseline against which the observed distance should be evaluated. Moreover, this upward bias originates from the finite-sample problem in the data examined. Although the relatively large samples and low-dimensional response spaces in existing studies may mitigate its magnitude, this bias can become more severe in smaller subgroup comparisons or more complex and sparse data structures. As political scientists broaden their EMD applications, the discipline faces a growing need to recognize this risk of overestimation.

Therefore, this paper cautions against interpreting standard bootstrap uncertainty bounds as a correction for the upward bias of empirical EMD. To address this problem, it proposes a permutation-based null calibration framework. By estimating the expected magnitude of the EMD under the null hypothesis of no group difference, this non-parametric approach establishes a finite-sample null baseline that the observed EMD can be interpreted relative to. Although the proposed approach does not yield an unbiased point estimate of the underlying population distance, it allows researchers to distinguish true distributional divergence from the positive EMD noise under the null, while not requiring any unrealistic directional or distributional shape assumptions.

This paper first discusses the EMD and its upward bias. Next, it discusses the limitations of standard empirical bootstrapping. Then, it discusses the problem in the context of sparse data structures using conjoint experiments as an illustrative case. Subsequently, it explains the proposed calibration method. Finally, it presents simulation results and compares the proposed method with other alternatives before offering some concluding remarks.

\section{EMD and Upward Bias}
Consider two empirical probability distributions, $\hat{P}_n$ and $\hat{Q}_m$, sampled from true underlying population distributions $P$ and $Q$ over a metric space $\mathcal{X}$ equipped with a ground metric $d(\cdot, \cdot)$. These empirical distributions are typically represented as probability vectors $\mathbf{p}$ and $\mathbf{q}$ over a finite support of $K$ distinct points within the metric space.\footnote{Regardless of whether $K$ denotes a fixed number of histogram bins (as in the traditional measures which require binning) or the union of unique continuous coordinates observed across the samples, the data always reduces to discrete probability vectors $\mathbf{p}$ and $\mathbf{q}$ over a finite support.} The EMD between these two empirical profiles is defined as the solution to the following linear optimization problem: 
\begin{equation}
    W_1(\mathbf{p}, \mathbf{q}) = \min_{\mathbf{T} \in \Pi(\mathbf{p}, \mathbf{q})} \sum_{i=1}^K \sum_{j=1}^K T_{ij} d(\mathbf{x}_i, \mathbf{x}_j)
\end{equation}
where $\Pi(\mathbf{p}, \mathbf{q})$ represents the set of all valid joint distributions whose marginal distributions satisfy $\sum_{j=1}^K T_{ij} = p_i$ and $\sum_{i=1}^K T_{ij} = q_j$ (the total probability masses mapped to coordinates $\mathbf{x}_i$ and $\mathbf{x}_j$, respectively), and $T_{ij}$ represents the mass transported. The distance is bounded by a non-negativity constraint ($W_1(\mathbf{p}, \mathbf{q}) \geq 0$). It yields a value of exactly zero if and only if the two empirical distributions are identical ($\mathbf{p} = \mathbf{q}$) on the given support.

Under the null hypothesis that the two samples are drawn from identical underlying populations ($H_0:P=Q$), the true population distance is zero ($W_1(P, Q) = 0$). However, in practice, due to finite sampling error, the realized empirical vectors are almost never identical ($\mathbf{p} \neq \mathbf{q}$). Since $W_1$ is a strictly non-negative functional, these random fluctuations cannot cancel out in the same way as signed differences in means. Instead, any empirical discrepancy maps into a positive transport cost whenever a mass is moved across the support space. Therefore, under the null, the expected empirical distance is generally positive, even when $P=Q$:
\begin{equation}
    \mathbb{E}\!\left[W_1(\hat{P}_n,\hat{Q}_m)\right] > 0
\end{equation}
where the severity of this problem depends on the geometry and dimensionality $D$ of the metric space, as I will demonstrate later. See the Appendix for the Kantorovich–Rubinstein dual formulation and the full proof of this strict positivity \autocite{berger_wasserstein_2009}. 

Notably, instead of implying that the estimator is inconsistent, this suggests that the problem is a finite-sample one. As $n,m \to \infty$, the empirical measures converge to their population counterparts, and the finite-sample bias decreases toward zero under standard regularity conditions. However, in practice, political science applications may involve subgroup applications that slice the data into smaller pieces with a smaller effective sample size. The data structures may also be sparse, especially when the scales are fine-grained. The positive null distance may then remain large enough to be mistaken for substantive group divergence. 

If the raw empirical EMD is interpreted as direct evidence of substantive divergence without a finite-sample null reference, this upward bias may increase the risk of false-positive conclusions on the existence of meaningful divergence (Type I error). This inferential problem can also propagate downstream when uncalibrated EMD estimates are subsequently used as dependent variables, predictors, or components of derived quantities in statistical models. Treating standard bootstrap uncertainty bounds as a substitute for a true null calibration baseline could further exacerbate this inferential problem. 

\section{Limitations of Standard Bootstrapping}
Existing literature has adopted bootstrapping to calculate uncertainty bounds around observed sample EMD values \autocites{lupu_new_2017, cummins_threat_2026, broderstad_trustees_2025, belschner_representation_2026}. Bootstrapping is often treated as a safety net for non-normal estimators to conduct statistical inference \autocite{harden_bootstrap_2011, jacoby_bootstrap_2014, kertzer_elite_2026, bimber_social_2022, mooney_bootstrapping_1993}. These applications primarily target the sampling variance of the empirical plug-in estimator. This approach quantifies its descriptive stability conditional on the observed data, rather than aiming to correct its absolute magnitude. An inferential problem can occur when these uncertainty bounds are treated as a direct test of whether an observed distance is substantively distinct from zero. Under the null, the population EMD equals zero and lies on the boundary of its parameter space ($W_1 \geq 0$), where its sampling behavior is non-standard. Conventional bootstrap intervals may therefore fail to provide valid null calibration for testing distributional equality.

In the standard bootstrapping approach, a bootstrap sample is generated by resampling with replacement from the already finite empirical distributions, $\mathbf{p}$ and $\mathbf{q}$. The resulting bootstrap distributions, $\mathbf{p}^*$ and $\mathbf{q}^*$, would approximate the sampling behaviour of the empirical plug-in estimator conditional on the observed data. When the original empirical EMD is already shifted upward by finite-sample noise, the bootstrap distribution will tend to reproduce this shifted baseline and estimate variability around it. The corresponding bootstrap approximation is therefore:
\begin{equation}
\mathbb{E}^*\!\left[W_1(\mathbf{p}^*, \mathbf{q}^*)\right] - W_1(\mathbf{p}, \mathbf{q}) \approx \mathbb{E}\!\left[W_1(\hat{P}_n, \hat{Q}_m)\right] - W_1(P, Q)
\end{equation}
Since $W_1$ is non-negative and sensitive to discrepancies in empirical mass allocation, the bootstrap distribution may remain centered away from zero even when $W_1(P, Q)=0$. 

This raises caution about applying standard bootstrap confidence interval (CI) procedures to the EMD without accounting for its non-standard asymptotic properties. Since the statistic is non-negative and can be concentrated near the boundary of zero, bias-correction procedures may behave poorly unless their finite-sample properties are explicitly checked. When faced with a known systematic upward shift, a common first line of defense is to apply a standard non-parametric bootstrap bias correction, such as the first-order bootstrap bias-corrected estimator \autocite{efronJackknifeBootstrapOther1982, efronIntroductionBootstrap1993}: 
\begin{equation}
    W_{\text{BC}} = 2W_1(\mathbf{p}, \mathbf{q}) - \mathbb{E}^*\!\left[W_1(\mathbf{p}^*, \mathbf{q}^*)\right]
\end{equation} 
However, the bootstrap expectation $\mathbb{E}^*\!\left[W_1(\mathbf{p}^*, \mathbf{q}^*)\right]$ is driven by the already noisy empirical baseline. Whenever it exceeds $W_1(\mathbf{p}, \mathbf{q})$, the bias-corrected estimate falls below the original empirical value ($W_{\text{BC}} < W_1(\mathbf{p}, \mathbf{q})$). If this discrepancy is sufficiently large that $\mathbb{E}^*\!\left[W_1(\mathbf{p}^*, \mathbf{q}^*)\right] > 2W_1(\mathbf{p}, \mathbf{q})$, $W_{\text{BC}}$ becomes negative ($W_{\text{BC}} < 0$). The resulting estimate therefore falls outside the
non-negative parameter space of the EMD metric.\footnote{This limitation could generalize broadly to other classical resampling and bias-correction frameworks applied near a hard non-negativity boundary.} 

Admittedly, this boundary limitation could be addressed by abandoning the standard bootstrap on the pooled distributions to construct conventional CIs in favor of modifying the sampling target. An alternative approach is to repeatedly partition the baseline sample into two equal-sized subsamples $\frac{N_{\text{baseline}}}{2}$ and calculate their distance to map purely stochastic baseline variation as a benchmark \autocite{cummins_threat_2026}. While this effectively characterizes within-group stability, in most congruence research, the sample size often fluctuates across experimental conditions or filtering criteria.\footnote{In the alternative case of LLM-human comparisons (mirroring), sample-size symmetry might exist provided non-refusal and perfect synthetic response validity.} The actual evaluation tests calculate the distance between the full baseline sample ($N_{\text{baseline}}$) and a comparison group of size $M$, and the finite-sample null distribution of empirical EMD depends on the sample sizes of both groups. There is no guarantee of symmetry, so this sample-size mismatch limits the baseline’s ability to serve as a precise null calibration. See the Appendix for a detailed proof.

\section{Sparse Data Structures}
The risk of upward bias discussed above increases when moving from low-dimensional continuous data to more sparse data structures. As Appendix A documents in detail, existing applications predominantly employ one-dimensional EMD over relatively small response spaces. This predominance limits one important source of finite-sample upward bias that will be discussed below, namely increasing dimensionality $D$. Nonetheless, the EMD offers the advantage of being capable of measuring congruence across multiple dimensions. Observationally, such multi-dimensional preference structures could arise in joint issue spaces that evaluate multiple domains concurrently.\footnote{When \textcite{lupu_new_2017} introduce the EMD, they run a multi-dimensional EMD analysis over a $D=7$ joint issue space. They observe that the negative effect of economic development on congruence appears only in the seven-dimensional issue space and not in the one-dimensional case. Raw high-dimensional EMD bias could have possibly inflated the result.} Experimentally, multi-dimensional preferences are frequently measured through factorial design, especially conjoint experiments \autocite{hainmuellerCausalInferenceConjoint2014, bansakConjointSurveyExperiments2021}. Such designs can create a sparsity problem for the EMD because the number of possible profiles grows rapidly with the number of attributes and levels.\footnote{This concerns estimators evaluating the joint distribution of profile choices via the EMD, instead of comparisons of marginal attribute-level choices that are often assessed using the Hellinger distance or Pearson's correlation of marginal choice frequencies.} 

Despite not having been widely adopted in the context of conjoint experiments, the EMD is useful when researchers want to compare the overall distribution of preferred or selected profiles across groups beyond estimating the treatment effect. As the EMD compares full profile distributions while taking into account the distance between profiles (two profiles that differ on one attribute are treated as closer than two profiles that differ on many attributes), it could be substantially meaningful. Meanwhile, given the inherent sparse data structures in conjoint experiments, their structural properties provide a useful illustration of how sparsity can exacerbate this upward shift.

Consider a standard conjoint design consisting of $A$ attributes, where each attribute $a$ has $L_a$ levels.\footnote{The attributes do not necessarily share the same number of levels in actual conjoint studies.} The total number of unique profile combinations, which defines the support space $\mathcal{S}$ with a total cardinality of $K = |\mathcal{S}|$, is:
\begin{equation}
\label{eq:support}
K = |\mathcal{S}| = \prod_{a=1}^A L_a
\end{equation} 
For a basic experimental design with six attributes, each featuring three levels, the support space $\mathcal{S}$ contains $K = 3^6 = 729$ unique profile configurations. When a researcher seeks to estimate choice shares or joint preference distributions across this full profile space, the target empirical distribution must allocate probability mass across several hundred, if not thousands, distinct categories. When the data points are distributed across them, the resulting empirical preference vectors ($\mathbf{p}$ and $\mathbf{q}$) are fundamentally sparse.\footnote{The key quantity of interest in conjoint experiments, average marginal component effects (AMCEs), reduces high-dimensional choice data by aggregating across marginal attribute levels ($A$), implicitly assuming that higher-order interaction effects among attributes can be smoothed or averaged out \autocite{hainmuellerCausalInferenceConjoint2014}. In contrast, evaluating congruence via the EMD over full conjoint profile distributions makes no dimensional-reduction or interaction-smoothing assumptions. Instead, it evaluates probability mass across the full joint support space $\mathcal{S}$.} This creates many empty or weakly populated cells when $K$ is large relative to $n$. 

As \textcite{fournier_rate_2015} show, for empirical measures on a $D$-dimensional continuous metric space, the expected error of the empirical EMD estimator deteriorates as $D$ increases. For $W_1$, the expected empirical error is characterized by the rates:
\begin{equation}
\label{eq:rate}
\mathbb{E}\left[W_1(\hat{P}_n, P)\right] =
  \begin{cases}
  O\left(n^{-1/2}\right) & \text{if } D = 1 \\
  O\left(n^{-1/2} \log(1+n)\right) & \text{if } D = 2 \\
  O\left(n^{-1/D}\right) & \text{if } D > 2
  \end{cases}
\end{equation}
Although a conjoint profile design constitutes a discrete support space, similar problems arise when the finite support points $\mathbf{x}_i$ are embedded as coordinates within a high-dimensional metric space $\mathcal{X}$ equipped with a ground distance. As the geometric or effective dimension $D$ scales with the complexity of the profile vector, convergence becomes more demanding. In applied settings, this asymptotic slowdown can leave subgroup empirical distributions poorly approximated. Since unregularized EMD evaluates exact mass displacement, it may treat sampling zeros as meaningful absences of preference. If a subgroup contains observations in a rare profile cell while another does not, the optimization routine may interpret this as distributional divergence and assign transport cost across the ground metric $d(\mathbf{x}_i, \mathbf{x}_j)$, thus inflating the group distance.

\section{Permutation-Based Null Calibration}
To resolve the inferential limitations of standard bootstrapping in the presence of sample-size asymmetry and strict non-negativity boundaries, this paper proposes a permutation-based null calibration. Instead of treating the observed empirical EMD as a direct estimate of the true population distance, this approach estimates how large the EMD would be expected to be under the null hypothesis of no group difference. The observed EMD can then be interpreted relative to this finite-sample null baseline.

The primary inferential object in this framework is the permutation distribution of the EMD statistic under the null hypothesis of exchangeable group labels. Suppose we observe two samples with empirical probability vectors $\mathbf{p}$ and $\mathbf{q}$ and sample sizes $n$ and $m$, respectively. We pool the observations into a single empirical reference distribution, which serves as the estimated common distribution under the null.\footnote{This pooling mechanism implies a subtle change to the conceptual nature of the null hypothesis itself. Instead of measuring intra-population heterogeneity (testing whether a comparison group falls within the bounds of expected population variance), the permutation baseline tests inter-population exchangeability (testing whether the two groups share an identical underlying distribution). Since an unregularized EMD treats every empty cell in a sparse conjoint matrix as a geometric penalty, the assessment of the alignment of the comparison group is ultimately contingent on this choice of baseline calibration.} We then repeatedly shuffle the group labels, partitioning the pooled data into permuted empirical vectors $\tilde{\mathbf{p}}$ and $\tilde{\mathbf{q}}$ while preserving the original sample sizes $n$ and $m$. For each permutation $b=1,\ldots,B_{\mathrm{perm}}$, the shuffled labels define two mock empirical vectors, $\tilde{\mathbf{p}}^{(b)}$ and $\tilde{\mathbf{q}}^{(b)}$, from which we compute a permuted EMD:
\begin{equation}
    \tilde{W}_1^{(b)} = W_1\!\left(\tilde{\mathbf{p}}^{(b)},\tilde{\mathbf{q}}^{(b)}\right)
\end{equation}
Repeating this procedure over $B_{\mathrm{perm}}$ permutations approximates the EMD distribution that would arise under the null hypothesis solely due to finite-sample fluctuation, sample-size imbalance, and sparsity in the observed support. The observed statistic can then be compared against this permutation-derived null distribution. A permutation-based $p$-value (with a finite-sample correction) is:
\begin{equation}
\label{eq:p}
    p = \frac{ 1+\sum_{b=1}^{B_{\mathrm{perm}}} \mathbf{1}\!\left\{\tilde{W}_1^{(b)} \geq W_1^{\text{obs}}\right\} }{B_{\mathrm{perm}}+1}
\end{equation}
where $W_1^{\text{obs}} = W_1(\mathbf{p},\mathbf{q})$ is the observed statistic. This $p$-value estimates the probability of observing an EMD at least as large as the empirical EMD if the two groups were in fact drawn from the same underlying distribution. 

The validity of this $p$-value follows from exchangeability, and detailed proofs can be found in the Appendix. Conditional on the pooled sample, all $\binom{n+m}{n}$ assignments of $n$ observations to one group and $m$ observations to the other group are equally likely under $H_0:P=Q$. Therefore, the observed label assignment is uniformly distributed among the possible assignments, and the rank of the observed statistic among the permutation statistics is uniform up to ties. It follows that the permutation test controls the Type I error rate at level $\alpha$, such that under the null:
\begin{equation}
    \label{eq:p-alpha}
    \Pr_{H_0}(p\leq \alpha)\leq \alpha
\end{equation}

This result also clarifies the scope of the proposed calibration, which creates a conditional null reference distribution. The permutation distribution describes the distribution of $W_1(\tilde{\mathbf{p}},\tilde{\mathbf{q}})$ that would arise if the realized pooled support, empirical locations, and pooled frequencies were held fixed, and only the group labels were allowed to vary under the null. The procedure does not calibrate against the additional uncertainty involved in drawing the pooled support from the population. 

This distinction matters especially in sparse or heavy-tailed settings ($n \ll K$). If the realized pooled sample contains rare or isolated support points, these points may exert disproportionate influence on both the observed empirical distance and the permutation reference distribution. For example, a low-mass support point may lie far from the main region of both distributions. If it appears in only one sample, it may account for a large share of the empirical transport cost. Permutation can assess whether the realized group assignment of such a point is unusual under exchangeability, but it cannot remove the influence of the point itself from the finite-sample realization being conditioned on.

Therefore, the proposed framework should not be interpreted as the full sampling distribution of $W_1(\hat P_n,\hat Q_m)$ or an unconditional debiasing estimator of the population EMD. When $P\neq Q$, the sampling behavior of the empirical EMD depends on the unknown population distributions, group-specific variances, tail behavior, support geometry, and dependence structure of the data. By construction, label permutation removes these group-specific features by imposing exchangeability. Researchers should leverage support diagnostics and sensitivity checks, such as the ratio of observations to support size and the number of low-frequency support points, to assess whether the result might be driven by a small number of isolated support points. If so, the calibrated excess distance should be interpreted as unstable even if the permutation test remains valid conditional on the realized pooled support. 

\subsection{Supplementary Descriptive Quantity}
Beyond the $p$-value, the permutation distribution can also be used as a calibration baseline for a supplementary descriptive quantity. The permutation-calibrated excess EMD summarizes how far the observed EMD exceeds the expected finite-sample null distance:
\begin{equation}
      W_{\text{excess}}(\mathbf{p},\mathbf{q}) = \max\left(0,\, W_1^{\text{obs}} - \bar{W}_{1,\text{null}} \right)
\end{equation}
where $\bar{W}_{1,\text{null}} =\frac{1}{B_{\mathrm{perm}}} \sum_{b=1}^{B_{\mathrm{perm}}}  \tilde{W}_1^{(b)}$ denotes the average EMD across the permutation distribution. This statistic should not be interpreted as an unbiased estimator of the population EMD $W_1(P,Q)$. Rather, it is a descriptive measure of the extent to which the observed empirical EMD exceeds the level expected under the finite-sample null due to sampling noise, sample-size imbalance, and sparsity. The permutation $p$-value should therefore remain the primary inferential result.

\subsection{Confidence Intervals}
While bootstrapping often comes with the purpose of obtaining CIs and quantifying uncertainty \autocite{zrimsek_quantifying_2026}, CIs for the adjusted quantity require additional assumptions. The finite-sample validity of the permutation test does not automatically imply that conventional two-sided CIs for adjusted EMD quantities are valid. Clamping the adjusted metric at zero via the $\max(0, \cdot)$ operator creates a non-differentiable boundary constraint under the null. This could result in a point mass at exactly zero that invalidates standard asymptotic assumptions for traditional two-sided CIs. See the Appendix for supplementary discussions of how CIs may be handled in this framework, if readers hope to compute them.

\section{Simulation}
To demonstrate the aforementioned logic, this paper performs four sets of Monte Carlo simulations.\footnote{The replication materials can be found on the Political Analysis Harvard Dataverse at \url{https://doi.org/10.7910/DVN/ZF60N9} \autocite{hung_replication_2026}.} All EMDs are computed as optimal transport problems between empirical probability measures using \textsf{POT} \autocite{flamary_pot_2021}. For continuous point-cloud simulations, observations are treated as equally weighted empirical measures over support points in a compact subset of the metric space, specifically $\mathcal{X} = [0,1]^D$. In this special case, the optimal transport problem is equivalent to an optimal matching problem between equally weighted observations. For discrete conjoint simulations, empirical distributions are represented as frequency vectors over the full conjoint profile support. The ground cost matrix is computed over normalized profile coordinates, so that the empirical profile frequencies map directly onto probability vectors over the shared support space.

\subsection{Asymptotic Convergence and Upward Bias}
The first simulation assesses the rate of empirical convergence and finite-sample upward bias under the true null. It repeatedly generates two independent samples of size $n \in \{20, 50, 100, 250, 500, 1000, 2000\}$ from identical, continuous uniform distributions ($P = Q$) across dimensions $D \in \{1, 2, 3\}$. Each observation is assigned equal probability mass, and EMDs are computed over 1,000 simulations.

\begin{figure}
    \centering
    \includegraphics[width=\linewidth]{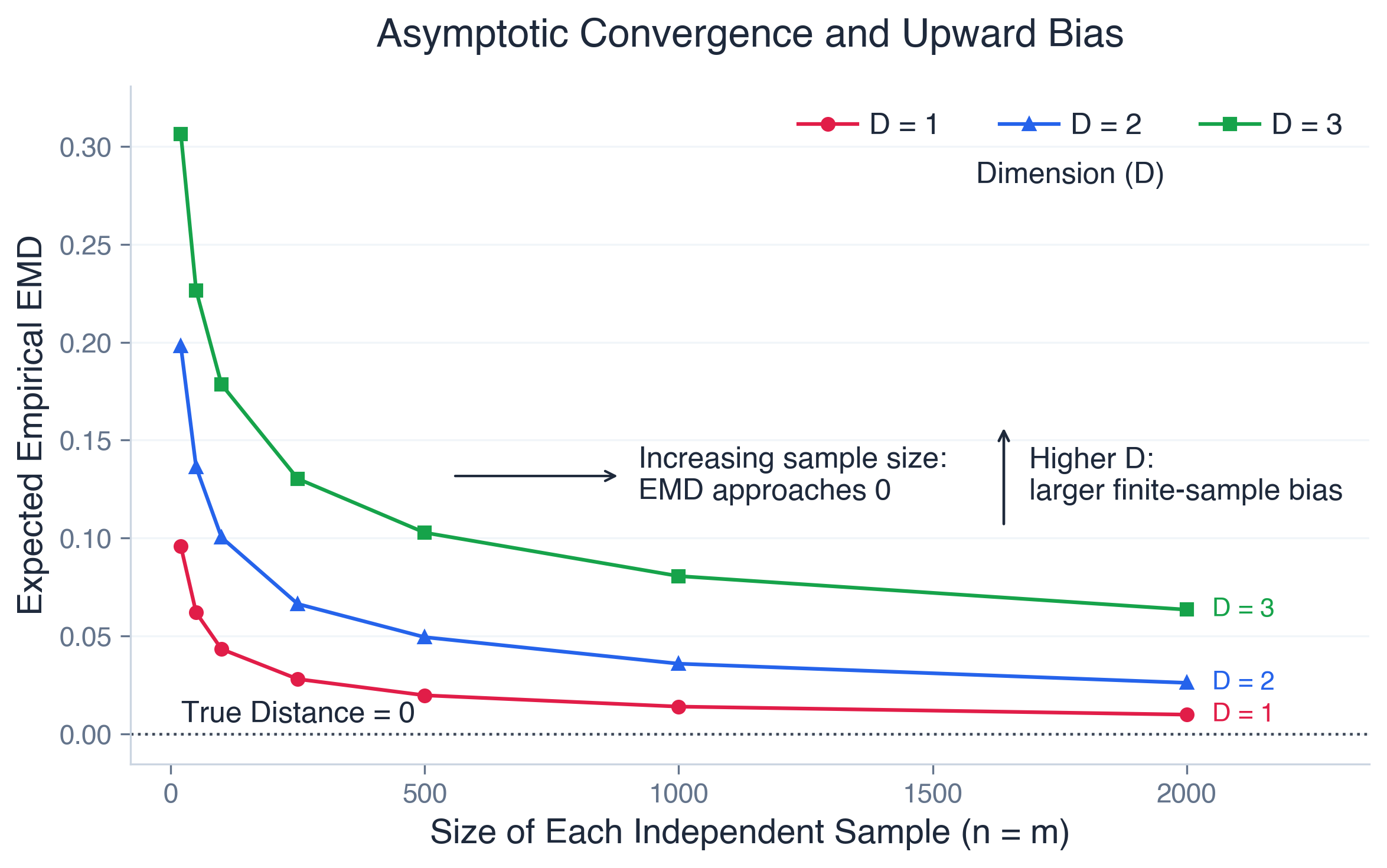}
    \caption{Asymptotic Convergence and Upward Bias}
    \label{fig:1}
\end{figure}

As Figure~\ref{fig:1} shows, random sampling variation produces a strictly positive empirical distance even when the two samples are drawn from the same underlying distribution. This upward shift decreases as $n \to \infty$ and increases with $D$. This illustrates why raw empirical EMDs may be substantively misleading in finite-sample settings, especially when researchers compare distributions over multi-dimensional preference spaces.

\subsection{Empirical Bootstrap Variance Inflation}
The second simulation evaluates the internal stability and variance of the empirical EMD distance using bootstrap resampling. It draws two continuous uniform samples in dimension $D \in \{1, 2, 3\}$ at a small sample size ($n=50$) and a larger sample size ($n=500$). An initial empirical sample pair is drawn to establish a fixed observed distance. 2000 bootstrap iterations are then performed to resample observations with replacement within each group to construct bootstrap EMD estimates.

\begin{figure}
    \centering
    \includegraphics[width=\linewidth]{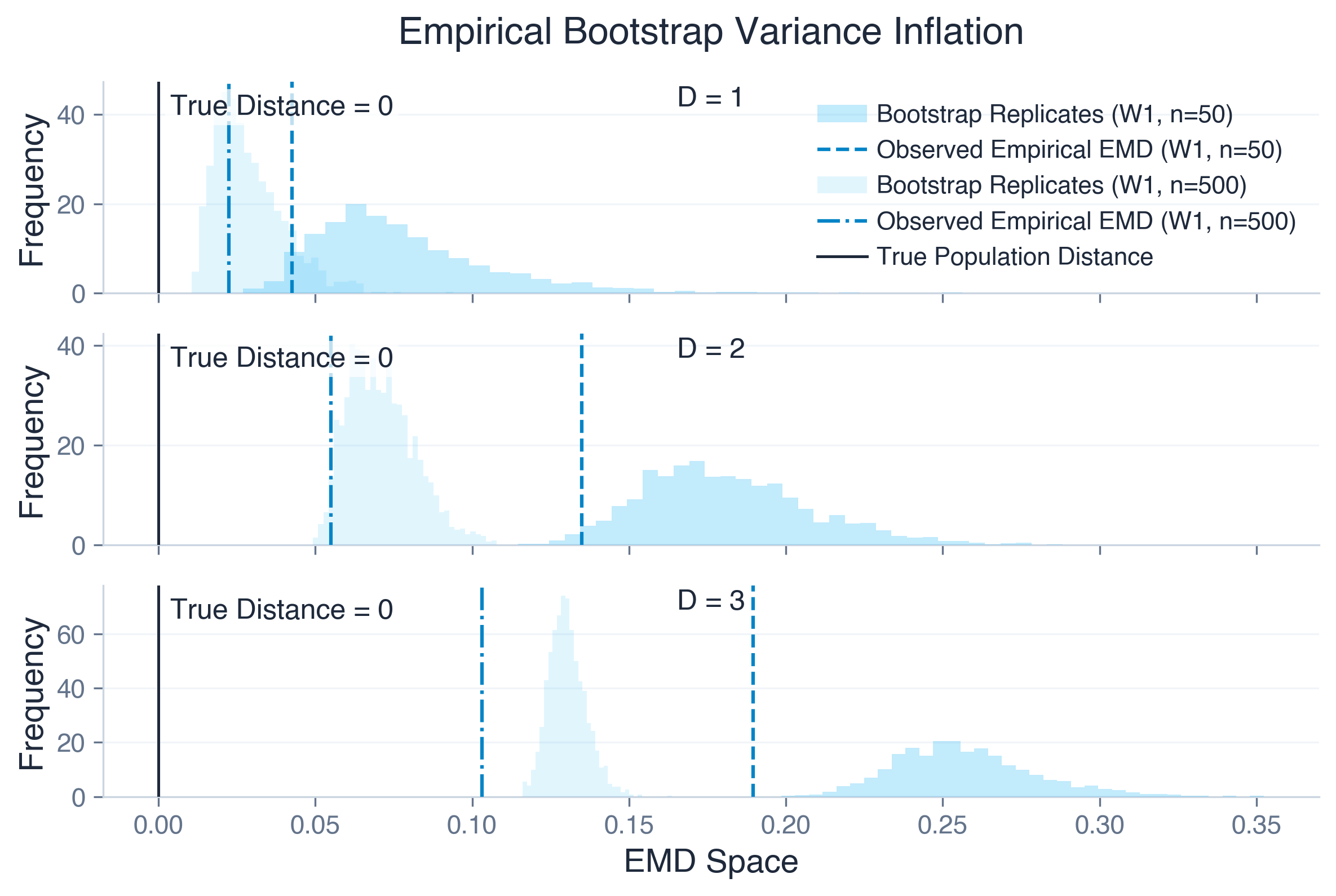}
    \caption{Empirical Bootstrap Variance Inflation}
    \label{fig:2}
\end{figure}

As Figure~\ref{fig:2} shows, the resulting bootstrap distributions are shifted away from zero rather than recovering the true null distance because the bootstrap samples are drawn from already noisy empirical measures. The bootstrap procedure therefore reproduces the finite-sample discrepancies present in the original empirical allocation, although the bias shrinks as $n$ increases and expands as $D$ increases as in Simulation 1. This further illustrates that standard empirical bootstrapping should not be interpreted as a correction for the upward bias of raw empirical EMDs.

\subsection{Distance Inflation in Sparse Conjoint Spaces}
The third simulation contextualizes these dimensionality problems within an applied setting exhibiting sparse data structures. It mimics the discrete support spaces typical of multi-attribute conjoint experiments. The simulation varies subgroup sample size across $n \in \{50, 100, 200, 500\}$ and generates profiles based on an attribute grid where each attribute has $L = 3$ possible levels.\footnote{These subgroup sample sizes should not be confused with the total number of participants in an entire study, i.e., the aggregated sum of participants across all concerned subgroups.} Increasing the number of attributes from $A=1$ to $A=6$ expands the support space $K = L^A$ exponentially from $3$ to $729$.\footnote{This also increases the geometric dimensionality, so the simulation is best interpreted as illustrating the combined finite-sample consequences of increasing dimensionality and the resulting support sparsity. The independent role of extreme support sparsity is demonstrated analytically in the Appendix through a simple discrete categorical example.} Under the true null, empirical profile counts are drawn independently for two groups over this shared support, and EMDs are computed using the discrete-support optimal transport procedure described above.

\begin{figure}
    \centering
    \includegraphics[width=\linewidth]{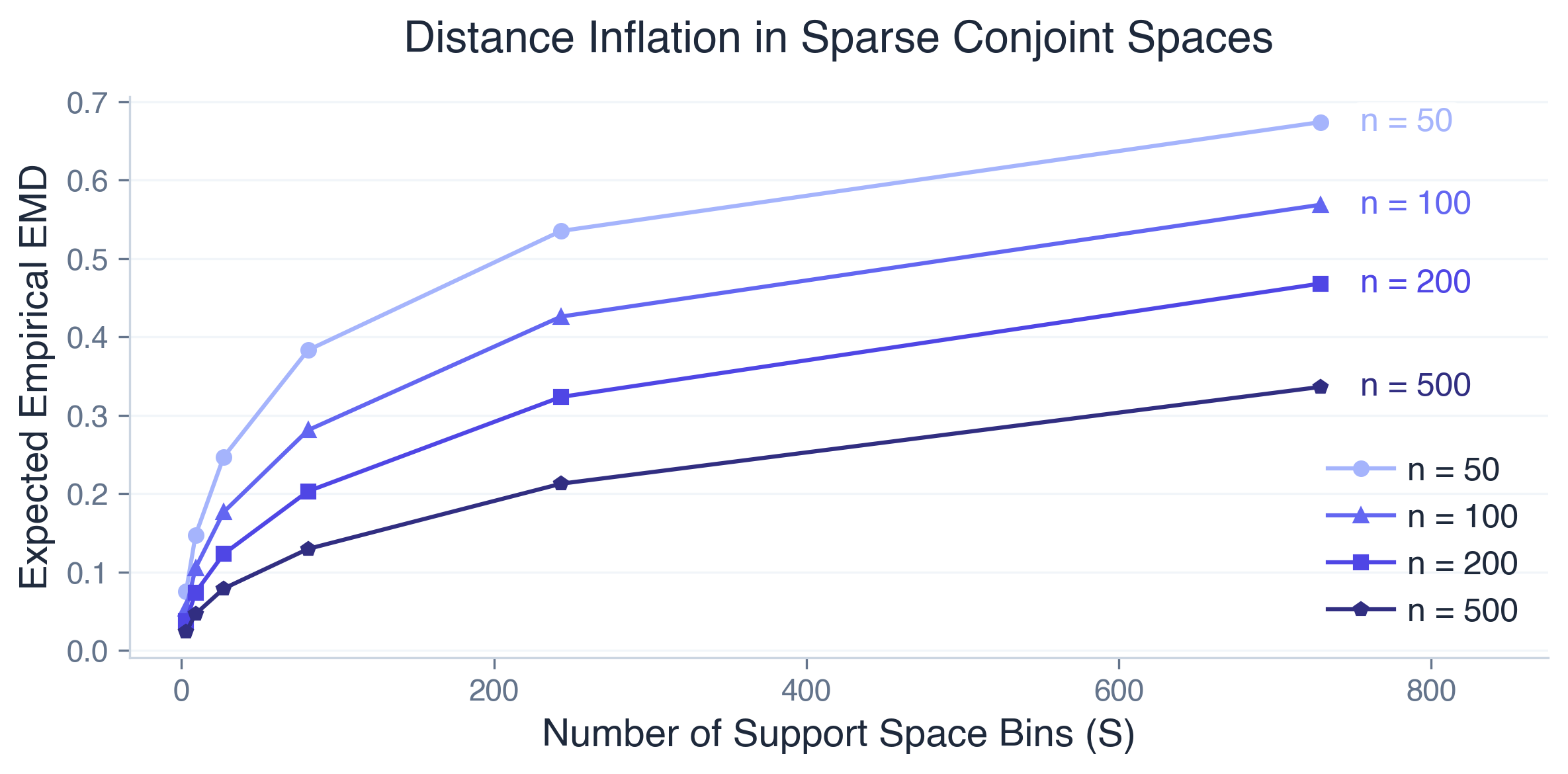}
    \caption{Distance Inflation in Sparse Conjoint Spaces}
    \label{fig:3}
\end{figure}

As Figure~\ref{fig:3} shows, the expected EMD increases as the conjoint support becomes more complex and fragmented. When the subgroup sample size remains small relative to this support, the empirical distributions become increasingly sparse and contain many sampling zeros, which arise even when the two groups are drawn from the same underlying distribution. However, unregularized optimal transport treats differences in empirical mass allocation as transport-relevant discrepancies. While increasing the sample size reduces the magnitude of this upward shift, it does not eliminate the underlying sparsity problem when the support cardinality $K$ remains large relative to the number of observations $n$. 

\subsection{Hypothesis Testing Calibration and Power}
The final simulation evaluates the proposed permutation-based hypothesis testing framework across continuous and sparse discrete settings. For each simulation run, I compute the observed EMD and apply the proposed permutation procedure using $B_{\mathrm{perm}}=500$ random label permutations to obtain the calibrated $p$-value.\footnote{In actual conjoint experiments, since each respondent has multiple tasks, the resampling or permuting should be conducted at the respondent level instead of the profile level to account for within-respondent clustering. However, given the simulations' objective to evaluate the EMD's finite-sample behavior under sparse conditions, I simplify the setup and hold the profile generation strictly independent and identically distributed (i.i.d.).} The first setting uses continuous point-cloud data. The baseline case draws observations from uniform distributions over $[0,1]^D$, while the baseline alternatives shift the second group to $[0.1,1.1]^D$, where $D=2$.\footnote{The continuous calibration simulations fix $D=2$ because the effect of dimensionality on finite-sample EMD behavior is examined in previous simulations. The validity of the permutation test under the null hypothesis follows from exchangeability and does not depend on a particular dimensionality. Extending the power analysis across dimensions would also complicate interpretation, since holding coordinate-level shifts fixed changes the overall multivariate separation as $D$ increases.} To assess whether calibration depends on this simple uniform design, I also consider a skewed null in which both groups are drawn independently from $P=Q=\mathrm{Beta}(2,5)$ in each dimension. Power is then evaluated across three families of alternatives.\footnote{These scenarios are not intended to be least-favorable or minimally detectable alternatives, but rather as graded robustness checks.} The first consists of uniform location shifts of varying magnitudes, where one group is drawn from $[0,1]^D$ and the other from $[\delta,1+\delta]^D$ for $\delta \in \{0.05, 0.075, 0.10\}$. The second consists of dispersion alternatives, where the two groups have the same mean but different spreads. I compare distributions, including $\mathrm{Beta}(6,6)$ versus $\mathrm{Beta}(4,4)$, $\mathrm{Beta}(7,7)$ versus $\mathrm{Beta}(4,4)$, and $\mathrm{Beta}(8,8)$ versus $\mathrm{Beta}(3,3)$. The third consists of bimodal polarization alternatives, where both groups are mixtures of two bounded normal components, but the second group places more probability mass toward the extremes of the support. They compare $((0.48,0.52))$ versus $((0.45,0.55))$, $((0.47,0.53))$ versus $((0.43,0.57))$, and $((0.44,0.56))$ versus $((0.38,0.62))$, respectively, with the component standard deviation fixed at $0.10$ throughout. These polarization designs evaluate whether the procedure can detect differences in distributional shape rather than merely location. 

The second setting uses a sparse conjoint histogram setting. It mimics a pooled choice-based conjoint design where a sample of 100 respondents each completes 10 tasks featuring $A = 6$ attributes and $L = 3$ levels per attribute. This yields a pooled sample size of $n = 1,000$ profile observations per group distributed over a discrete multi-attribute support space.\footnote{A pooled sample size of $n = 1,000$ ensures that the total number of observations exceeds the total cardinality of the support space ($L^A = 3^6 = 729$). This avoids the low statistical power problem, where severe under-sampling relative to the bin count flattens out the geometric signals needed to detect alternative distributions.} Under the null condition, both groups are drawn uniformly from the full conjoint profile support, whereas under the alternative condition, the second group is drawn from a distribution that places a systematic preference weighting on higher attribute levels. 

To assess sensitivity to the strength and scope of this preference shift, I also generate alternatives using exponential level weights proportional to $\exp (\beta l)$, where $l \in \{1, \dots, L\}$ indicates the attribute level. The parameter $\beta \in \{0.1,0.2,0.5,1.0\}$ controls the strength of the preference shift, where smaller values produce alternatives closer to the uniform null distribution. I compute the weights using a stable softmax function in which the raw exponent values $z_i = \beta l_i$ are shifted by subtracting their maximum value prior to exponentiation, such that the probability for each level is evaluated as: 
\begin{equation}
P(z_i) = \frac{\exp(z_i - \max_j(z_j))}{\sum_j \exp(z_j - \max_j(z_j))}
\end{equation} 
I evaluate these four values of $\beta$ and repeat the same procedure used in the main simulation.\footnote{For $\beta=0.1$, the weights are $[0.3006, 0.3322, 0.3672]$; for $\beta=0.2$, they are $[0.2693, 0.3289, 0.4018]$; for $\beta=0.5$, they are $[0.1863, 0.3072, 0.5065]$; and for $\beta=1.0$, they are $[0.0900, 0.2447, 0.6652]$.} Moreover, to assess whether the high statistical power in the main simulation and the previous one is driven by a globally coordinated shift across all attributes, I consider a more conservative localized alternative in which only one attribute is shifted while the remaining attributes are sampled uniformly.

\begin{figure}
    \centering
    \includegraphics[width=\linewidth]{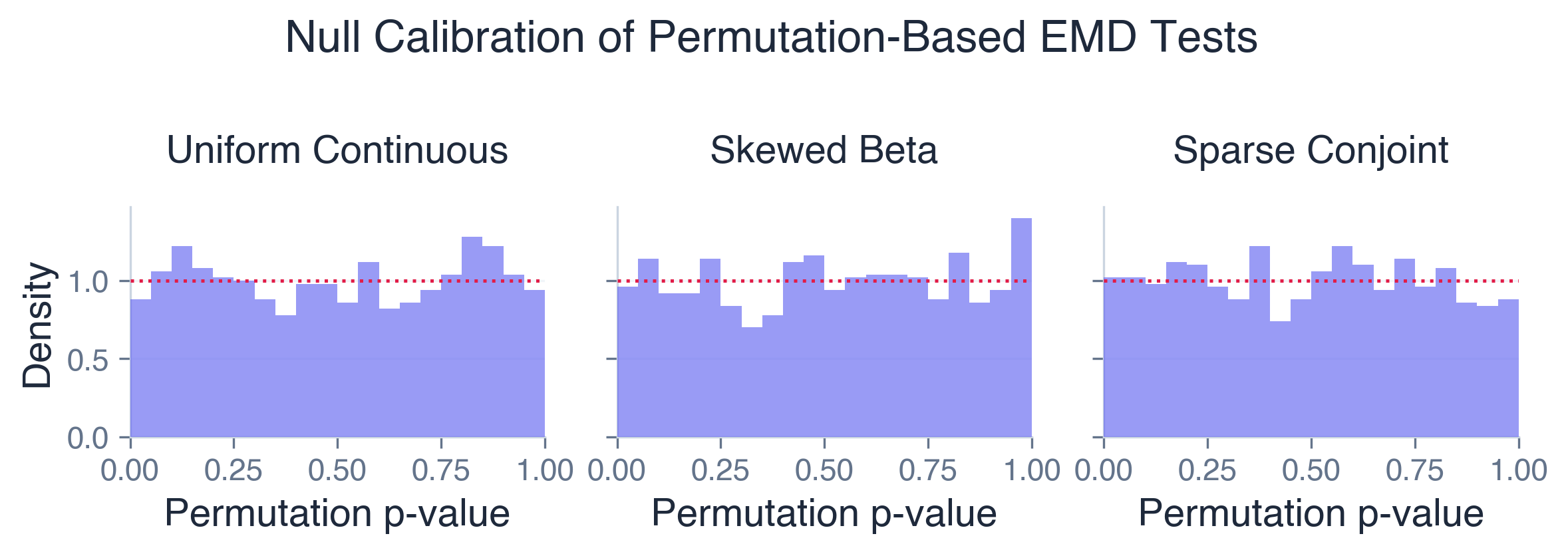}
    \caption{Null Calibration of Permutation-Based EMD Tests}
    \label{fig:4}
\end{figure}

As Figure~\ref{fig:4} and Table~\ref{tab:power} show, the null $p$-values are approximately uniform across the scenarios, with the rejection rates under the null hypothesis remaining close to the nominal 5\% level. This indicates that the permutation procedure properly calibrates the test under the null. Under the alternative condition, the results show the expected signal-dependent graded pattern across the settings. Subtle distributional differences are detected less frequently, whereas power increases as the separation becomes more pronounced. Likewise, the sparse conjoint results show a similar pattern. Global preference shifts are detected reliably once the shift is moderate, whereas weak local one-attribute shifts remain difficult to detect. This suggests that the strong results in the baseline sparse conjoint simulation are not merely a result of one particular preference-weighting scheme. Overall, the results indicate that the procedure controls Type I error while retaining sensitivity to sufficiently pronounced differences.

\begin{table}[!htbp]
\centering
\footnotesize
\caption{Size and Power of Permutation-Based EMD Tests}
\label{tab:power}
\begin{tabular}{llc}
\hline
Setting & Scenario & Rejection rate \\
\hline
\multicolumn{3}{l}{\textit{Null scenarios: Type I error}} \\
Continuous & Uniform null & 0.044 \\
Continuous & Skewed Beta null, $\mathrm{Beta}(2,5)$ & 0.048 \\
Sparse conjoint & Uniform null & 0.051 \\
\hline
\multicolumn{3}{l}{\textit{Continuous alternatives: Power}} \\
Continuous & Location shift, $\delta=0.05$ & 0.252 \\
Continuous & Location shift, $\delta=0.075$ & 0.553 \\
Continuous & Location shift, $\delta=0.10$ & 0.829 \\
Continuous & Dispersion, $\mathrm{Beta}(6,6)$ vs. $\mathrm{Beta}(4,4)$ & 0.241 \\
Continuous & Dispersion, $\mathrm{Beta}(7,7)$ vs. $\mathrm{Beta}(4,4)$ & 0.480 \\
Continuous & Dispersion, $\mathrm{Beta}(8,8)$ vs. $\mathrm{Beta}(3,3)$ & 0.990 \\
Continuous & Polarization, $(0.48,0.52)$ vs. $(0.45,0.55)$ & 0.078 \\
Continuous & Polarization, $(0.47,0.53)$ vs. $(0.43,0.57)$ & 0.193 \\
Continuous & Polarization, $(0.44,0.56)$ vs. $(0.38,0.62)$ & 0.784 \\
\hline
\multicolumn{3}{l}{\textit{Sparse conjoint alternatives: Power}} \\
Sparse conjoint & Global shift, $\beta=0.1$ & 0.526 \\
Sparse conjoint & Global shift, $\beta=0.2$ & 1.000 \\
Sparse conjoint & Global shift, $\beta=0.5$ & 1.000 \\
Sparse conjoint & Global shift, $\beta=1.0$ & 1.000 \\
Sparse conjoint & Local one-attribute shift, $\beta=0.1$ & 0.109 \\
Sparse conjoint & Local one-attribute shift, $\beta=0.2$ & 0.354 \\
Sparse conjoint & Local one-attribute shift, $\beta=0.5$ & 1.000 \\
Sparse conjoint & Local one-attribute shift, $\beta=1.0$ & 1.000 \\
\hline
\end{tabular}

\begin{flushleft}
\footnotesize
Note: Under null scenarios, the rejection rate estimates Type I error. Under alternative scenarios, the rejection rate estimates statistical power.
\end{flushleft}
\end{table}

\section{Comparison with Relevant Alternatives}
Up to this point, this paper has demonstrated the utility and robustness of the proposed calibration approach. Before wrapping up, there are some relevant alternatives worth discussion. First, the Sinkhorn divergence is often used as an alternative to the EMD \autocites{cuturi_sinkhorn_2013, feydy_interpolating_2018}. It addresses the upward bias by modifying the standard entropic regularization framework. The debiased formulation explicitly corrects for it by subtracting the self-transportation costs of each individual distribution.\footnote{Note that here the upward bias is induced by regularization instead of sampling.} Given its inherent advantage in computational speed, it could be practically useful, especially in the case of machine learning optimization. However, there is a trade-off in political science, where researchers also care about the interpretability of the distance. Unlike the proposed framework, the Sinkhorn divergence does not offer an exact, unregularized physical interpretation of moving mass across the coordinate space $\mathcal{X}$, since the transportation plan is smoothed.

Second, \textcite{papp_centered_2022} introduce a pair of centered plug-in estimators based on linear combinations. Their approach is computationally efficient but focuses on continuous distributions with connected support and negligible boundaries. These estimators are developed for the squared $W_2$ case, which penalizes transport distance quadratically instead of linearly as in the classic EMD ($W_1$), and require overdispersion between distributions to remain conservative.\footnote{In contrast, the proposed framework can be applied to both $W_1$ and $W_2$ metrics. See the Appendix for detailed discussions.} In the one-dimensional case, the centered plug-in estimator can be applied to both continuous and discrete distributions because one-dimensional optimal transport can be computed directly using the inverse cumulative distribution functions of the measures. However, this does not apply to the multi-dimensional case, but the EMD is appreciated for its ability to evaluate multi-dimensional divergence. Also, discrete distributions are common in political science research, while polarization and other problems like boundary effects are prevalent and could render the assumptions invalid. In contrast, although both measures preserve the unregularized scale and exact geometry of the space that the Sinkhorn alternative cannot offer, the proposed calibration approach is non-parametric. It does not require any directional or distributional shape assumptions. In short, unless researchers can safely assume one distribution is inherently more spread out than the other and both distributions satisfy the continuous requirements, the proposed calibration approach will be more robust while also retaining the physical interpretability.

\section{Conclusion}
While the non-negativity problem associated with distance metrics is relatively general, as optimal transport exploits geometric discrepancies across the support and empirical convergence deteriorates in higher-dimensional spaces, extra caution should be exercised in the case of the EMD. Even if the use case fortunately involves low dimensionality and fairly large sample sizes that can help reduce the upward bias, acknowledging and addressing this bias (if needed) would add extra robustness to the result. As the EMD is applied to more data structures to study distribution similarity, this finite-sample distortion will become increasingly important for political science research.

This paper has illustrated that the standard bootstrapping framework is not by itself a solution for resolving the upward bias caused by finite-sample noise facing political scientists. Since bootstrapping resamples from an already noisy empirical allocation, it tends to reproduce the finite-sample baseline rather than recover the true population distance. To address this inferential problem, this paper proposes an easily implementable permutation-calibration framework, which helps determine whether an observed empirical distance exceeds what would be expected from finite-sample noise, sparse support, and the study’s design structure under the null. As a non-parametric approach, it especially benefits the political science discipline, where the data do not necessarily fulfill various assumptions that alternative estimators demand, such as continuity. 

More broadly, as the positive bias problem applies to many other distance metrics, the logic of the proposed calibration can also be extended to create a finite-sample null reference for a statistic whose raw magnitude is otherwise easy to over-interpret. Also, although this paper is motivated by the comparison of preference distributions central in political science research, the methodological argument is not limited to preferences. It applies more generally to empirical comparisons of two probability distributions defined on a common metric space, provided that the ground distance between support points is substantively meaningful.

Conceptually, while generating a null distribution by shuffling labels is a familiar tool in non-parametric statistics \autocite{mooney_bootstrapping_1993}, the proposed framework treats the permutation distribution beyond a passive testing mechanism. In standard empirical applications, permutation frameworks are utilized strictly as an intermediate step to extract a $p$-value for a binary rejection decision. Here, the distribution is considered a foundational inferential object that maps and calibrates against the inherent finite-sample upward bias. It enables researchers to exploit the full geometric profile of the sample space under the null hypothesis rather than compressing it into a single localized probability threshold. This way, the framework helps place empirical evaluations of congruence and preference distributions on a more methodologically defensible foundation.

\section{Bibliography}
\printbibliography[heading=none]

\appendix
\counterwithin{figure}{section}
\counterwithin{table}{section}
\counterwithin{equation}{section}

\section{Appendix: Recent EMD Applications}
\label{sec:application}
Table~\ref{tab:application} shows a list of examples of recent political science applications of the EMD. These studies are primarily identified by examining the papers citing \textcite{lupu_new_2017} through CrossRef, while the LLM-related ones are identified through targeted searches and are included if they engage with political science-related topics or data. I also manually screen false positives that engage with the discussion of the metric or distance measures without directly applying the EMD.  

\begin{sidewaystable}[!htbp]
\centering
\tiny
\caption{Recent EMD Applications in Political Science}
\label{tab:application}
\begin{tabularx}{\textwidth}{
p{2.0cm} p{2.5cm} p{3.2cm} p{3.2cm} p{2.4cm} p{3.8cm} X}
\hline
Paper & Application &
Distributions compared &
Sample entering each EMD &
Dimension/Support &
Role in empirical framework &
Treatment of sampling noise \\
\hline

\textcite{lupu_new_2017}
& Mass-elite ideological and issue congruence
& CSES voters vs.\ seat-weighted parties; LAPOP citizens vs.\ PELA legislators
& Varies by country-wave; mass and elite sample sizes unequal
& 1D left-right ($K=11$); multi-dimensional survey items with a $D=7$ joint issue space
& Primary descriptive metric; Congruence measure used as DV in cross-national OLS regressions; Latin American application uses separate regressions of EMD on each covariate
& Acknowledges upward bias; recommends bootstrap uncertainty bounds \\

\textcite{devine_convergence_2021}
& European citizen-party-system congruence
& Eurobarometer/ESS voter distributions vs.\ expert-survey party positions
& Country-wave voter $n$ varies; party side is seat-weighted party distribution
& 1D, issue-by-issue; left-right harmonized to 0-10
& DV in country- and year-fixed-effects regressions testing effects of EU integration on party-system, parliament, and government congruence
& None; country jackknife used as regression robustness check \\

\textcite{helliesen_unequal_2023}
& Mass-elite climate-policy congruence
& NCP citizen vs.\ PER representative distributions, including subgroups
& Wave- and subgroup-specific ($1155$-$8445$ citizens and $1593$-$4053$ representatives)
& 1D ($K=7$)
& DV in two OLS regressions testing whether EMD differs across sociodemographic subgroups
& None; subgroup-size comparability considered \\

\textcite{sorace_does_2023}
& Systemic representation failure and populist voting
& Election-specific CSES voter distribution vs.\ seat-weighted non-populist party distribution
& Election-specific; $106105$ respondents across 64 elections in total
& 1D left-right
& Election-year-level EMD used as a robustness operationalization of sociotropic incongruence and entered as an IV in individual-level fixed-effects logistic regressions predicting populist vote choice
& None; EMD itself used as robustness measure \\

\textcite{carella_electoral_2024}
& Geographic unrepresentativeness (SURLI)
& MP birthplace distribution vs.\ gridded population distribution
& $>13000$ legislators across 62 democracies; population represented by grid weights
& 2D spatial target, approximated by four rotated 1D EMDs; country-specific grid support
& Component of composite index (SURLI = observed/null EMD), serving as DV in regressions
& 500 random-legislature null simulations; observed EMD divided by mean null EMD \\

\textcite{ball_human_2025}
& LLM persona-to-party mappings
& Model-implied party distribution vs.\ weighted GLES party distribution
& Aggregated personas and prompt variants
& Distribution over six parties ($K=6$)
& Benchmark distance to GLES voting distribution; also prompt-sensitivity proxy in regressions relating Wasserstein distance to normalized entropy
& None for finite-sample EMD bias \\

\textcite{cao_specializing_2025}
& LLM simulation of global survey-response distributions
& Model first-token probability distribution vs.\ country-level WVS/Pew response proportions
& Human reference $n>1000$ per WVS country; model side is a probability vector
& 1D response-option distribution; $K$ varies by question
& Evaluation metric (together with Jensen-Shannon Divergence) for model-human distributional fit
& None \\

\textcite{cummins_threat_2026}
& LLM silicon-sample fidelity
& Human vs.\ synthetic distributions of social-psychological and political scale scores
& 85 humans and approximately 85 synthetic observations per configuration
& 1D scores: Normalized $-9$-$9$ ($K=19$) and $6$-$36$ ($K=31$) scales
& Evaluation metric for the correspondence of response distributions in silicon samples with human data; Wasserstein-based distributional-fit score also serves as an outcome in exploratory univariate regressions of configuration features
& 2,000 repeated human split-half EMDs as reference benchmark \\

\textcite{fortunato_economic_2025}
& Cross-national voter-preference similarity
& Eurobarometer voter distribution in country A vs.\ country B
& $8554$ downstream dyad-years coming from 179 surveys of 30 European democracies between 1976 and 2022
& 1D ($K=10$)
& DV in dyadic time-series cross-sectional regressions on bilateral economic links
& None \\

\textcite{muriaas_attitudes_2025}
& Citizen-elite congruence in gender-quota support
& NCP citizen vs.\ PER representative support distributions
& Approximately 1500 citizens vs.\ $>4100$ representatives
& 1D ($K=7$)
& Ancillary descriptive congruence check (single overall EMD)
& None for EMD \\

\textcite{broderstad_trustees_2025}
& Citizen-elite referendum-legitimacy congruence
& Citizen vs.\ representative distributions by marginal treatment level
& 1568 citizens vs.\ 4321 representatives
& 1D ($K=2$)
& Complementary congruence metric across marginal treatment levels, used alongside marginal mean treatment analysis
& 1,000-iteration bootstrap confidence intervals \\

\textcite{belschner_representation_2026}
& Proxy representation and inclusion of unenfranchised groups
& National politician distributions vs.\ voter, underage youth, and migrant non-voter distributions
& Country-policy-group-specific; 1185 politicians vs.\ 27465 citizens (Approximately 400 underage youth and 1000 migrant non-voters) across 14 democracies
& 1D, issue-by-issue ($K=7$, 9 policy proposals)
& Comparative congruence metric in the mechanical-proxy-representation analysis, summarized across voter groups, policies, and countries
& Bootstrapped 95\% confidence intervals \\

\textcite{marzi_whom_2026}
& Mass-elite and elite-elite congruence and bias
& Party representative distributions vs.\ party voter and economic elite distributions
& Party-year-question-specific; 98 parties with $\geq 3$ surveyed representatives each; voter and economic-elite $n$ varies by country-wave; 2,610 party-year-issue triads
& 1D, issue-by-issue (37 items rescaled to $[0,1]$)
& DV as differenced metric ($\Delta W_1 = W_{1,\text{voters}} - W_{1,\text{elites}}$) in multilevel regressions
& None for finite-sample EMD bias; inference conducted downstream using difference-in-means tests and multilevel regressions \\

\textcite{sorace_europeanisation_2026}
& Cross-national policy-preference convergence
& EES voter distribution in country A vs.\ country B, by wave and issue
& 500 or 1000 respondents per country
& 1D per issue ($K=11$)
& DV in dyadic panel regressions testing preference convergence over time
& None \\

\hline
\end{tabularx}
\end{sidewaystable}

When \textcite{lupu_new_2017} introduce the EMD to political science, they discuss the use of the EMD in both one-dimensional and multi-dimensional settings, where non-negligible upward bias is especially plausible. Their Latin American application combines several survey items into a single multi-dimensional EMD comparing mass and elite preferences over a joint space of $D=7$. Although \textcite{lupu_new_2017} themselves recognize that sample-based EMD is likely to overstate population distance and recommend bootstrapping to quantify uncertainty, as discussed above, conventional bootstrap confidence intervals characterize sampling variability around the empirical statistic rather than directly estimating the positive EMD expected under a null hypothesis of identical population distributions. As the main paper also discusses, bootstrap uncertainty intervals have subsequently appeared in applied EMD research and often explicitly claimed as a mitigation of the upward bias problem.

Although \textcite{lupu_new_2017} introduce the EMD as a useful measure for multi-dimensional data, most subsequent applications concern one-dimensional data.\footnote{\textcite{lupu_new_2017} actually criticize that most congruence research focuses on the summary left–right ideological dimension, and attribute this problem to the limitations of the traditional measures. However, as the table shows, some applications continue to rely on left-right scales that political scientists have long been used to.} Note that \textcite{carella_electoral_2024} explicitly avoid running a full two-dimensional EMD due to computational limitations.\footnote{Besides the tendency to rely on one-dimensional ideological scales, this may explain the predominance of one-dimensional application of the EMD. \textcite{carella_electoral_2024} have noted that it may take weeks to compute two-dimensional EMDs for their study. This paper's simulations also take over weeks to run in full on a local machine, with the sparse conjoint case being the most demanding, even when parallel processing is used. While the EMD is known to be computationally demanding when $D>1$ and $K$ is large \autocite{pele_fast_2009, cuturi_sinkhorn_2013}, cloud computing could allow researchers to speed up EMD analysis by executing heavy, parallelizable optimal transport and distribution-matching tasks across massive remote clusters \autocite{chen_cloud_2018}. This could reduce execution time from days to minutes, although researchers may need to refactor their code according to the cloud platform's requirements.} They approximate the two-dimensional EMD by computing the one-dimensional EMD across multiple rotations and averaging them, although this also indicates a desire to compute the EMD beyond the one-dimensional use case and the authors also acknowledge that their approximation is not necessarily robust to other data. This predominance of one-dimensional applications limits one important source of finite-sample upward bias identified in this paper's simulations, namely increasing dimensionality $D$. Nevertheless, one-dimensional applications are not necessarily immune to substantively meaningful bias because the magnitude of the finite-sample noise floor also depends on factors, such as sample size, support size, and the balance between the distributions being compared.

The data structures used in these applications vary considerably in their susceptibility to such bias. Some studies compare relatively large samples over a small number of response categories. For example, \textcite{broderstad_trustees_2025} compare 1,568 citizens with 4,321 elected representatives using marginal one-dimensional EMDs for a binary outcome. Such a setting should generate considerably less finite-sample upward bias than a small-sample or high-dimensional application.\footnote{Note that the full joint space would have a $K=36$ instead of 2.} Similarly, \textcite{muriaas_attitudes_2025} compare approximately 1,500 citizens with more than 4,100 representatives on a seven-category scale, while \textcite{sorace_europeanisation_2026} generally compares country samples of 500 or 1,000 respondents on 0-10 policy scales. In these applications, a positive finite-sample EMD remains expected even when the underlying population distributions are identical, but whether this component is substantively important depends on its magnitude relative to the observed EMD. There might still exist borderline cases and the results may change when the proposed calibration is applied.

While the aggregated sample size of existing studies tends to be relatively large, existing applications can involve substantially smaller samples once EMD is calculated within countries or subgroups. For example, \textcite{belschner_representation_2026} draw on 27,465 citizens overall, but their EMD analysis separately compares politicians with underage youth and migrant non-voters by country and policy issue. These subgroups comprise only about 400 and 1,000 respondents, respectively, across all 14 countries. Thus, the effective samples entering individual EMD calculations can be considerably smaller than the headline sample size of an application suggests. \textcite{helliesen_unequal_2023} compares citizen and representative distributions not only overall but also within sociodemographic groups defined by gender, age, education, and region. This could reduce the subgroup size to only around 100, or even one- or two-digit numbers.\footnote{\textcite{helliesen_unequal_2023} explicitly notes that analyses with small $N$ may yield skewed results and therefore uses broader age and income categories in the main analysis to increase subgroup sizes and comparability. The smallest subgroup sample in their main analysis is slightly above 100. Nevertheless, the supplementary analyses report EMD estimates for substantially smaller subgroups.} In settings where these distributions are small or unevenly sized, the positive finite-sample component may consequently become more important.

Furthermore, the EMD may have downstream applications. For instance, \textcite{marzi_whom_2026} evaluate relative representational bias by taking the difference between two empirical distances ($\Delta W_1 = W_{1,\text{voters}} - W_{1,\text{elites}}$) and modeling this quantity across party-year-issue triads. Similarly, \textcite{helliesen_unequal_2023} use EMD itself as the dependent variable in regressions comparing sociodemographic groups, while \textcite{fortunato_economic_2025} and \textcite{sorace_europeanisation_2026} employ EMD-based congruence as the outcome in dyadic regression models. EMD can also enter on the right-hand side of a model, such as \textcite{sorace_does_2023}, which uses EMD-based sociotropic incongruence as an alternative explanatory variable in models of populist voting. These applications create an additional reason to distinguish empirical EMD from its finite-sample null component. A large number of downstream regression observations does not imply that the empirical distributions used to construct each observation are themselves large. Moreover, when a derived quantity such as $\Delta W_1$  subtracts two empirical EMDs constructed from comparison samples of different sizes, their positive finite-sample null distances need not be equal, so taking their difference does not necessarily eliminate finite-sample bias. More generally, when these uncalibrated distance metrics are subsequently fed into downstream regression models or difference-in-means tests, this finite-sample distortion risks propagating into the analysis. This could potentially bias estimated coefficients or increase Type I error rates, especially when the substantive difference between the component distances is small.

Ultimately, it should be clarified that Table~\ref{tab:application} does not imply that existing substantive conclusions based on EMD are necessarily invalid. Many applications use relatively large samples and low-dimensional response spaces, for which the upward bias may be small relative to the observed distances. A replication exercise, which is outside the scope of this paper, would be necessary to establish whether their results are highly sensitive to the use of permutation-based calibration. Still, the table demonstrates that several existing studies operate in settings where a non-negligible finite-sample component remains plausible, as well as imply the possibility of an exacerbation of upward bias issues when the EMD is extended to other use cases. In short, the proposed calibration remains helpful for adding robustness to research findings, even in cases where the upward bias might be relatively small.

\section{Appendix: Strict Positivity}
\subsection{Kantorovich-Rubinstein Dual Formulation}
The primal optimization problem of the EMD can be further illustrated by its Kantorovich-Rubinstein dual formulation \autocite{berger_wasserstein_2009}. Instead of searching for an optimal transport plan over pairs of support points, the framework identifies a 1-Lipschitz `witness' function that maximizes the expected difference between the two empirical vectors:
\begin{equation}
W_1(\mathbf{p}, \mathbf{q}) = \max_{\mathbf{f} \in \mathcal{F}} \sum_{i=1}^K f_i (p_i - q_i)
\end{equation}
where $\mathcal{F}$ denotes the set of all vectors $\mathbf{f} \in \mathbb{R}^K$ satisfying the Lipschitz continuity condition $f_i - f_j \leq d(\mathbf{x}_i, \mathbf{x}_j)$ for all pairs of coordinates $\mathbf{x}_i, \mathbf{x}_j \in \mathcal{X}$. This optimization framework inherently capitalizes on finite-sample noise, thus converting random fluctuations into a positive measured distance even when the true population levels are completely identical.

\subsection{Proof}
The strict positivity follows from the separation property of the EMD metric. Under $H_0:P=Q$, the population distance is $W_1(P,Q)=0$. However, the empirical measures $\hat{P}_n$ and $\hat{Q}_m$ are independently realized random measures. Since $W_1(\hat{P}_n,\hat{Q}_m)\geq 0$, and since $W_1(\hat{P}_n,\hat{Q}_m)=0$ if and only if $\hat{P}_n=\hat{Q}_m$, it follows that $\mathbb{E}_0[W_1(\hat{P}_n,\hat{Q}_m)]>0$ whenever $\Pr(\hat{P}_n\neq \hat{Q}_m)>0$. This condition holds in ordinary finite-sample settings unless the underlying distribution is degenerate. In continuous settings, two independently drawn empirical measures coincide with probability zero. On the other hand, in discrete settings, they may coincide with positive probability, but not absolutely unless the distribution is degenerate. Thus, even when the population distance is exactly zero, the empirical EMD has a positive finite-sample null expectation.

\subsection{Simple Binary Example of Strict Positivity}
\label{sec:eg-positivity}
After proving that the empirical EMD has a positive finite-sample null expectation, I proceed to illustrate the magnitude of this baseline with a simple binary example. 

Define a binary metric space $\mathcal{X}=\{0,1\}$ with $d(0,1)=1$, and suppose both groups are drawn from $\mathrm{Bernoulli}(\theta)$. If $X\sim \mathrm{Binomial}(n,\theta)$ and $Y\sim \mathrm{Binomial}(m,\theta)$ are independent counts of observations assigned to the point $1$ in the two groups, then the empirical probability masses at $1$ are $\hat{p}=\frac{X}{n}$ and $\hat{q}=\frac{Y}{m}$. In this binary case, the EMD between the two empirical distributions collapses exactly to the absolute difference in the mass assigned to the point $1$:
\begin{equation}
    W_1(\widehat P_n,\widehat Q_m) = \left|\frac{X}{n}-\frac{Y}{m}\right|
\end{equation}

The exact finite-sample null expectation is therefore:
\begin{equation}
    \mathbb{E}_0[W_1] = \sum_{x=0}^{n}\sum_{y=0}^{m} \left|\frac{x}{n}-\frac{y}{m}\right| \binom{n}{x}\theta^x(1-\theta)^{n-x} \binom{m}{y}\theta^y(1-\theta)^{m-y}
\end{equation}

For moderate to large $n$ and $m$, the Central Limit Theorem gives the standard normal approximation for the difference of two independent binomial proportions:
\begin{equation}
    \frac{X}{n}-\frac{Y}{m} \sim \mathcal{N}\left( 0, \theta(1-\theta) \left(\frac{1}{n}+\frac{1}{m}\right) \right)
\end{equation}

If $Z \sim \mathcal{N}(0, \sigma^2)$, then its mean absolute deviation is exactly $\mathbb{E}[|Z|] = \sigma \sqrt{\frac{2}{\pi}}$. Applying this result to the normal approximation above gives:
\begin{equation}
\label{eq:null-expect}
    \mathbb{E}_0[W_1] \approx \sqrt{\frac{2}{\pi}} \sqrt{ \theta(1-\theta) \left(\frac{1}{n}+\frac{1}{m}\right)}
\end{equation}

For example, when $\theta=0.5$ and $n=m=100$,
\begin{equation}
    \mathbb{E}_0[W_1]\approx 0.056
\end{equation}
Thus, even in the simplest possible two-category setting, a visibly positive empirical EMD arises purely from sampling noise.

\subsection{Simple Binary Example of Split-half Resampling Applications under Sample Size Asymmetry}
In the following, I use the split-half resampling approach employed by \textcite{cummins_threat_2026} to illustrate why a within-sample reference distribution does not generally provide an exact finite-sample null calibration when the sample sizes of the reference and target comparisons differ.

Under the split-half approach, the reference noise floor ($\text{Noise}_{\text{split-half}}$) is calculated using two equal subsamples of size $n_1 = n_2 = \frac{N_{\text{baseline}}}{2}$. Based on the finite-sample null expectation in discrete binary set-ups given by Equation~\ref{eq:null-expect}, the expected distance under this artificial null setup scales as:
\begin{equation}
    \mathbb{E}[\text{Noise}_{\text{split-half}}] \propto \sqrt{\frac{1}{\frac{N_{\text{baseline}}}{2}} + \frac{1}{\frac{N_{\text{baseline}}}{2}}} = \sqrt{\frac{4}{N_{\text{baseline}}}}
\end{equation}

However, when the sample sizes are asymmetric, the comparison evaluates the full baseline group ($n = N_{\text{baseline}}$) against an asymmetric comparison group of size $m = M$ (where $N_{\text{baseline}} \neq M$). The actual finite-sample null noise floor ($\text{Noise}_{\text{actual}}$) for this evaluation scales as:
\begin{equation}
    \mathbb{E}[\text{Noise}_{\text{actual}}] \propto \sqrt{\frac{1}{N_{\text{baseline}}} + \frac{1}{M}}
\end{equation}

The relative magnitude of the two noise floors is therefore:
\begin{equation}
\frac{\mathbb{E}[\text{Noise}_{\text{actual}}]}{\mathbb{E}[\text{Noise}_{\text{split-half}}]} = \sqrt{\frac{\frac{1}{N_{\text{baseline}}}+\frac{1}{M}}{\frac{4}{N_{\text{baseline}}}}} = \frac{1}{2}\sqrt{1+\frac{N_{\text{baseline}}}{M}}
\end{equation}

The split-half reference matches the finite-sample noise of the actual comparison only when $M=\frac{N_{\text{baseline}}}{3}$. When sample sizes are asymmetric, the directional bias of the split-half benchmark depends on the comparison group size $M$:
\begin{equation}
\frac{\mathbb{E}[\text{Noise}_{\text{actual}}]}{\mathbb{E}[\text{Noise}_{\text{split-half}}]}
\begin{cases}
<1 & \text{if } M>\frac{N_{\text{baseline}}}{3}\\
=1 & \text{if } M=\frac{N_{\text{baseline}}}{3}\\
>1 & \text{if } M<\frac{N_{\text{baseline}}}{3}
\end{cases}
\end{equation}
When $M>\frac{N_{\text{baseline}}}{3}$, the split-half procedure produces a higher noise floor than that associated with the actual comparison. If it were used as an inferential null distribution, this would yield a conservative calibration and reduce power. Conversely, when $M<\frac{N_{\text{baseline}}}{3}$, the split-half benchmark understates the finite-sample noise associated with the actual comparison and would yield an anti-conservative calibration if used for hypothesis testing.

In short, the empirical null distribution of EMD depends strictly on the exact pair of sample sizes $(n, m)$, sample-size mismatch between the split-half reference groups $\left(\frac{N_{\text{baseline}}}{2}, \frac{N_{\text{baseline}}}{2}\right)$ and the actual test groups $(N_{\text{baseline}}, M)$ invalidates the split-half benchmark as an exact null calibration.

\section{Appendix: Finite-Sample Validity Under Exchangeability}
The validity of the permutation $p$-value follows from the exchangeability of group labels under the null. Let $Z_1,\ldots,Z_N$ denote the pooled observations, where $N=n+m$. Under the null hypothesis $H_0:P=Q$, the observed group labels carry no systematic information about the distribution from which each observation was drawn. Conditional on the pooled sample, all label assignments that allocate $n$ observations to one group and $m$ observations to the other are therefore equally likely.

Let $\mathcal{G}$ denote the set of all $\binom{N}{n}$ possible label assignments preserving the original sample sizes. For each assignment $g\in\mathcal{G}$, define the test statistic as: 
\begin{equation}
    T(g)=W_1(\widehat P_g,\widehat Q_g)
\end{equation}
The observed assignment $g_{\mathrm{obs}}$ is one element of $\mathcal{G}$. Under the null, $g_{\mathrm{obs}}$ is uniformly distributed over $\mathcal{G}$. Therefore, the rank of $T(g_{\mathrm{obs}})$ among the permutation statistics $\{T(g):g\in\mathcal{G}\}$ is uniformly distributed up to ties.

It follows that under the null, the probability that the permutation $p$-value is less than or equal to the significance level $\alpha$ is no greater than $\alpha$: $$\Pr_{H_0}(p\leq \alpha)\leq \alpha$$ as Equation 9 in the main paper shows. Therefore, the test is finite-sample valid under exchangeability. This result does not require normality, differentiability of the Wasserstein functional, or an asymptotic approximation around the boundary $W_1=0$. In other words, the permutation baseline calibrates the observed statistic against the finite-sample distribution generated by the same support structure, sample sizes, and sparsity pattern as the original data, which is precisely why the permutation baseline is useful for empirical EMD.

\section{Appendix: Extreme Sparsity and Low Statistical Power}
Under extreme sparsity, when sample sizes are significantly mismatched with the cardinality of the support space ($n \ll K$), empirical distributions would collapse into isolated point masses scattered across a massive empty space. Under these conditions, the distance calculated by an unregularized linear program is entirely dominated by the geometric penalties of moving mass between isolated occupied cells across a largely unoccupied support. 

Shuffling the labels during a permutation test does not remove this sparsity. These tests condition on the pooled empirical support and repeatedly partition the data into two equally sparse resampled groups, whose baseline null distances are identically inflated. This causes the observed distance under a true alternative hypothesis to look entirely indistinguishable from the permuted null distributions. In other words, in extreme cases, the permutation test may have very low power because the null distribution of distances between randomly relabeled sparse samples is itself large and highly dispersed.

The actual likelihood of falling into this inferential zero-power trap depends heavily on the ratio between the total number of observations and the support cardinality $K$. When the unit of analysis is the total evaluated data points, the nominal ratio $\frac{N}{K}$ often exceeds one, which generally ensures that non-parametric permutation tests retain at least some capacity to reject the null. However, researchers should remain aware that real-world applications might be subject to severe effective sparsity. Even when profiles are randomized approximately uniformly, unless researchers pay extra care to the balance of utility of the profiles and their realism or behavioural plausibility \autocite{huber_importance_1996, rose_constructing_2009}, the empirical distribution of selected profiles may be highly concentrated in a small region of the support and the vast majority of the $K$ combinations will be weakly populated. Since unregularized EMD solves an exact mass-displacement problem across the entire space, it remains highly sensitive to these unpopulated tails. Even when the nominal count of data points appears robust, the estimator remains vulnerable to the geometric background noise of the empty cells, so the inflation of the baseline threshold will remain a persistent threat.

\subsection{Simple Discrete Categorical Example of Extreme Sparsity}
To illustrate this problem quantitatively, I provide a discrete categorical example below. This generalizes the logic outlined earlier in Appendix~\ref{sec:eg-positivity} to a discrete support space case of $K>2$ categories.

Consider two independent samples of sizes $n$ and $m$ drawn uniformly over a discrete support of $K$ points with the discrete ground metric $d(x_i, x_j) = 1$ for all $i \neq j$. Under the null hypothesis $H_0:P = Q = \text{Uniform}(K)$, each observation has probability $\frac{1}{K}$ of falling on any support point.

Let $X_k$ denote the number of observations from the first sample that fall on support point $x_k$, and let $Y_k$ denote the corresponding count for the second sample. Since each observation must fall into exactly one of the $K$ categories, the vectors of category counts follow multinomial distributions $(X_1,\ldots,X_K) \sim \mathrm{Multinomial} \left(n;\frac{1}{K},\ldots,\frac{1}{K}\right)$ and $(Y_1,\ldots,Y_K) \sim \mathrm{Multinomial} \left(m;\frac{1}{K},\ldots,\frac{1}{K}\right)$.

The empirical probability masses assigned to support point $x_k$ are therefore $\frac{X_k}{n}$ and $\frac{Y_k}{m}$. Under the discrete unit ground metric, the empirical $W_1$ distance is exactly the total variation distance between the two multinomial empirical distributions:
\begin{equation}
    \mathbb{E}_0 [W_1] = \frac{1}{2} \sum_{k=1}^K \mathbb{E} \left\vert{} \frac{X_k}{n} - \frac{Y_k}{m} \right\vert{}
\end{equation}
Since each of the $n$ observations either falls on support point $x_k$ with probability $\frac{1}{K}$ or does not, each individual count has a binomial marginal distribution, i.e., $X_k\sim\mathrm{Binomial}\left(n,\frac{1}{K}\right)$ and $Y_k\sim\mathrm{Binomial}\left(m,\frac{1}{K}\right)$.\footnote{The counts $X_1,\ldots,X_K$ are not independent, since they must sum to $n$. Likewise, $Y_1,\ldots,Y_K$ must sum to $m$.} Equivalently, $W_1$ in this case can be expressed as:
\begin{equation}
W_1 = 1-\sum_{k=1}^{K}\min\left(\frac{X_k}{n}, \frac{Y_k}{m}\right)
\end{equation}

When $n, m \ll K$, the fraction of empirical probability mass that overlaps tends to zero. Thus, the transport cost reaches its theoretical maximum bound:
\begin{equation}
\mathbb{E}_0[W_1] \to 1
\end{equation}

This implies that under extreme sparsity, the calculated distance can become dominated by finite-sample noise, while the permuted null distribution is similarly pushed toward near-maximum values. Consequently, the observed distance becomes increasingly difficult to distinguish from distances generated under the permutation null, leading to very low statistical power to detect true underlying differences. In the limiting case of complete empirical non-overlap, statistical power can approach zero.

\section{Appendix: Confidence Intervals}
In the following, I provide more details of how CIs can be used and interpreted in the proposed framework.

Although the clamped statistic has a boundary at zero and could create problems, researchers can bypass this issue by conducting statistical inference on the unclamped difference statistic: 
\begin{equation}
    W_{\text{diff}}(\mathbf{p}, \mathbf{q}) = W_1(\mathbf{p}, \mathbf{q}) - \mathbb{E}_{\text{null}}[W_1(\tilde{\mathbf{p}}, \tilde{\mathbf{q}})]
\end{equation} 
Unlike $W_{\text{excess}}$, this statistic is allowed to take negative values when the observed groups are more similar than expected under random group-label assignment. It helps researchers understand the relative position of the observed distance to the finite-sample null baseline. However, instead of assuming that $W_{\text{diff}}$ follows a standard continuous, symmetric, or approximately normal distribution, researchers should evaluate its behavior empirically. In sparse categorical settings, the permutation distribution may be discrete and irregular.

For this reason, the permutation-based $p$-value should remain the primary inferential object. Interval estimates for adjusted EMD quantities should be reported only as supplementary diagnostics unless their coverage properties are evaluated in the relevant design setting. When researchers wish to report uncertainty for the calibrated magnitude, a conservative alternative is to report one-sided bounds rather than conventional two-sided intervals. For example, an upper confidence bound can be constructed from the empirical distribution of an appropriate resampled or permuted version of the unclamped statistic:
\begin{equation}
    U_{1-\alpha} = \inf  \left\{u :P^*\!\left(W_{\text{diff}}^* \leq u\right) \geq 1-\alpha \right\}
\end{equation}
where $W_{\text{diff}}^*$ is computed by applying the same calibration procedure to each resampled dataset and $P^*$ denotes the resulting empirical resampling distribution used for the diagnostic interval. Such bounds should be interpreted cautiously as summarizing the uncertainty in the calibrated empirical quantity, but not necessarily the uncertainty in an unbiased estimate of the population EMD.

\section{Appendix: Extension}
While the standard EMD optimizes linear transport costs ($W_1$), this paper's framework extends naturally to the quadratic case, the 2-Wasserstein distance ($W_2$). The $W_2$ metric squares the ground distances within the optimization routine and takes the square root of the minimized total cost:
\begin{equation}
W_2(\mathbf{p}, \mathbf{q}) = \left( \min_{\mathbf{T} \in \Pi(\mathbf{p}, \mathbf{q})} \sum_{i=1}^K \sum_{j=1}^K T_{ij} [d(\mathbf{x}_i, \mathbf{x}_j)]^2 \right)^{1/2}
\end{equation}
Substantively, while $W_1$ reflects the minimum average shift required to align two distributions, $W_2$ acts as a root-mean-square deviation that disproportionately penalizes mass transported across greater distances in the preference space. 

Similar to $W_1$, $W_2$ metrics are vulnerable to systematic upward bias. In the case of bootstrapping, the resulting uncertainty bounds around $W_2$ metrics also face a distortion problem. This is especially likely when squared ground distances amplify the influence of extreme, low-probability support points. In the case of sparse data structures, this issue is even more severe. This, however, implies that the permutation-calibrated excess EMD introduced can be even more beneficial, as it can be easily generalized for any chosen order $r \in \{1, 2\}$:
\begin{equation}
W_{r,\text{excess}}(\mathbf{p}, \mathbf{q}) = \max\left(0, W_r(\mathbf{p}, \mathbf{q}) - \mathbb{E}_{\text{null}}[W_r(\tilde{\mathbf{p}}, \tilde{\mathbf{q}})] \right)
\end{equation}

\section{Appendix: Extension to the Broader Family of Wasserstein Metrics}
\begin{figure}
    \centering
    \includegraphics[width=\linewidth]{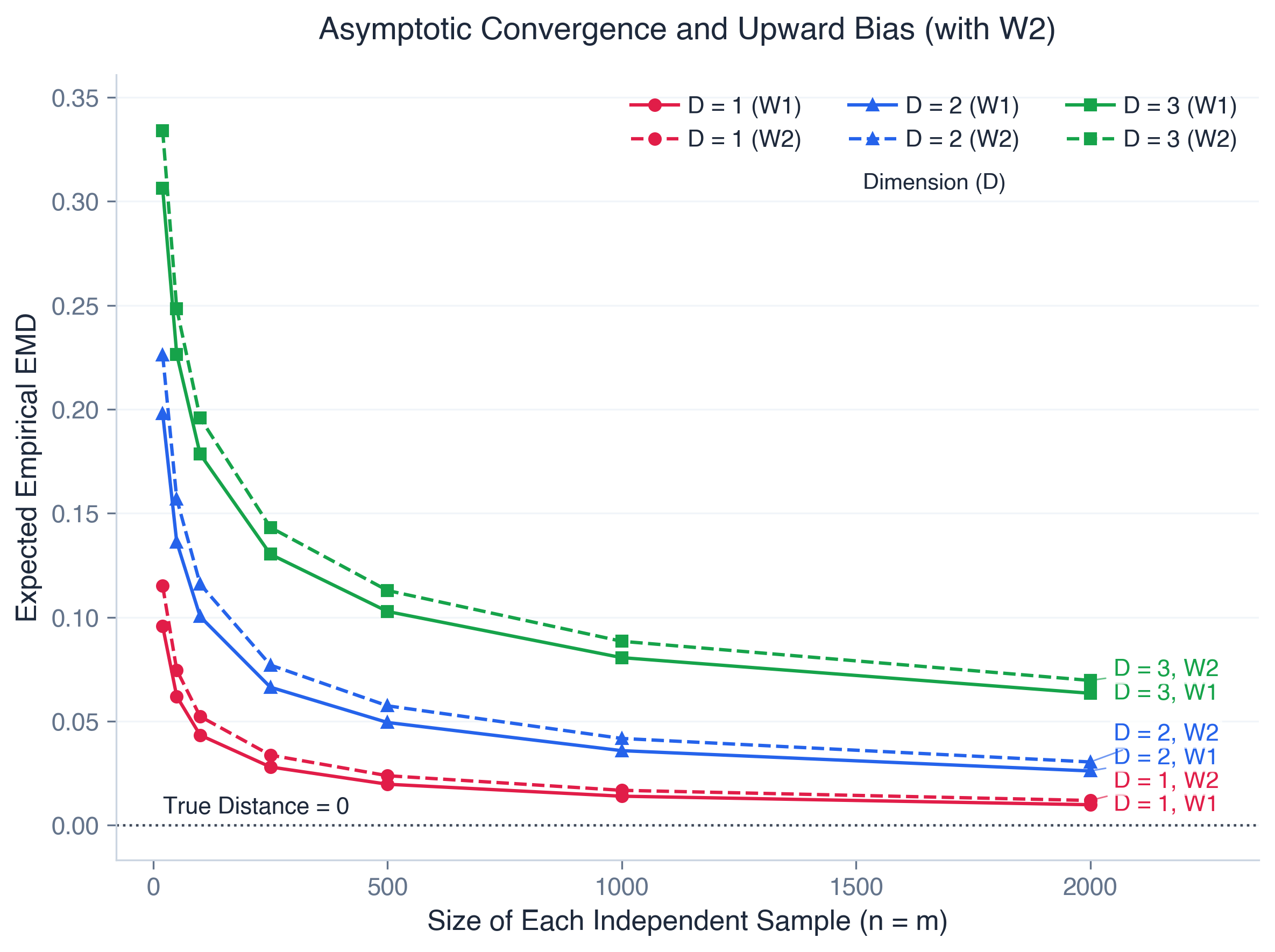}
    \caption{Asymptotic Convergence and Upward Bias (with $W_2$)}
    \label{fig:1-w2}
\end{figure}

\begin{figure}
    \centering
    \includegraphics[width=\linewidth]{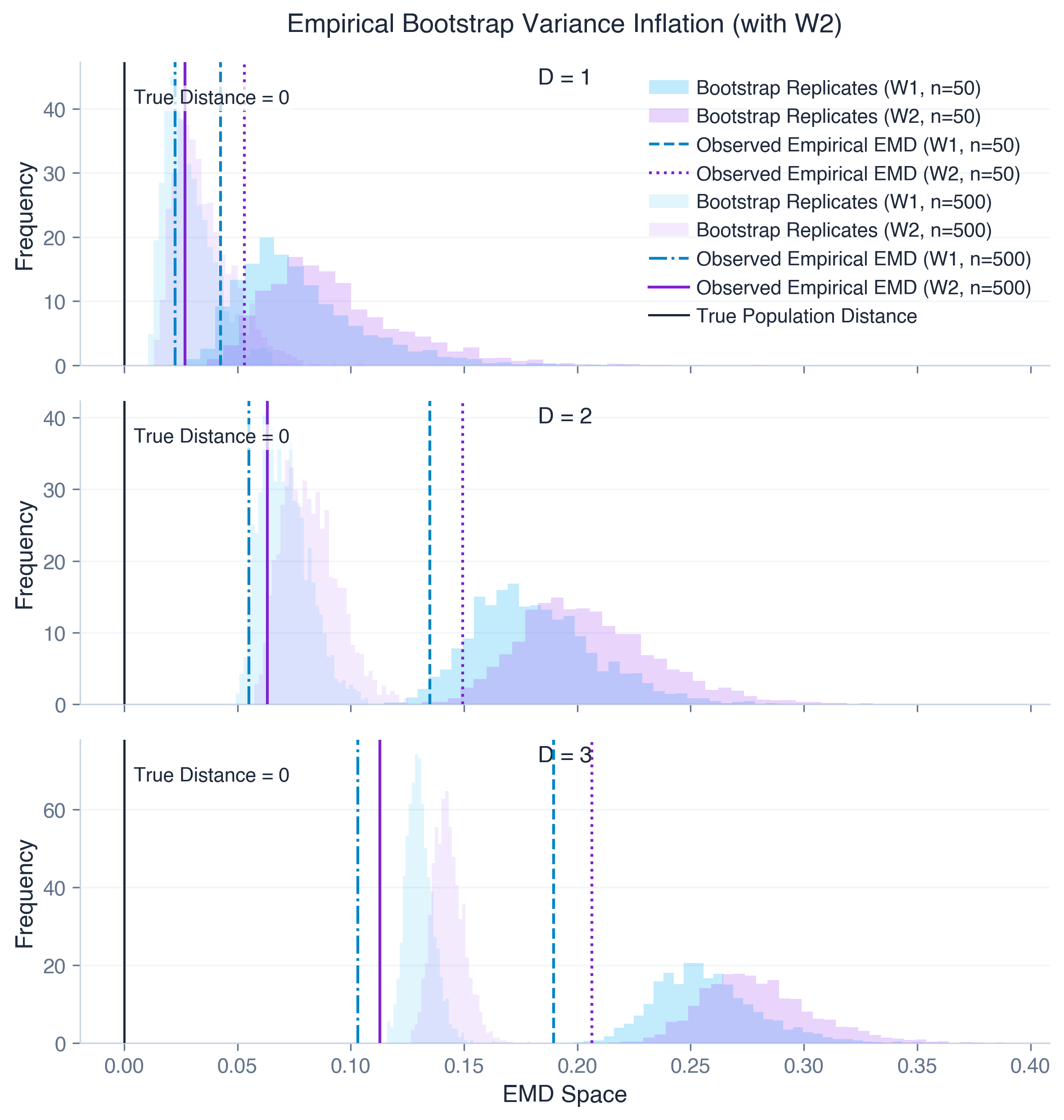}
    \caption{Empirical Bootstrap Variance Inflation (with $W_2$)}
    \label{fig:2-w2}
\end{figure}

\begin{figure}
    \centering
    \includegraphics[width=\linewidth]{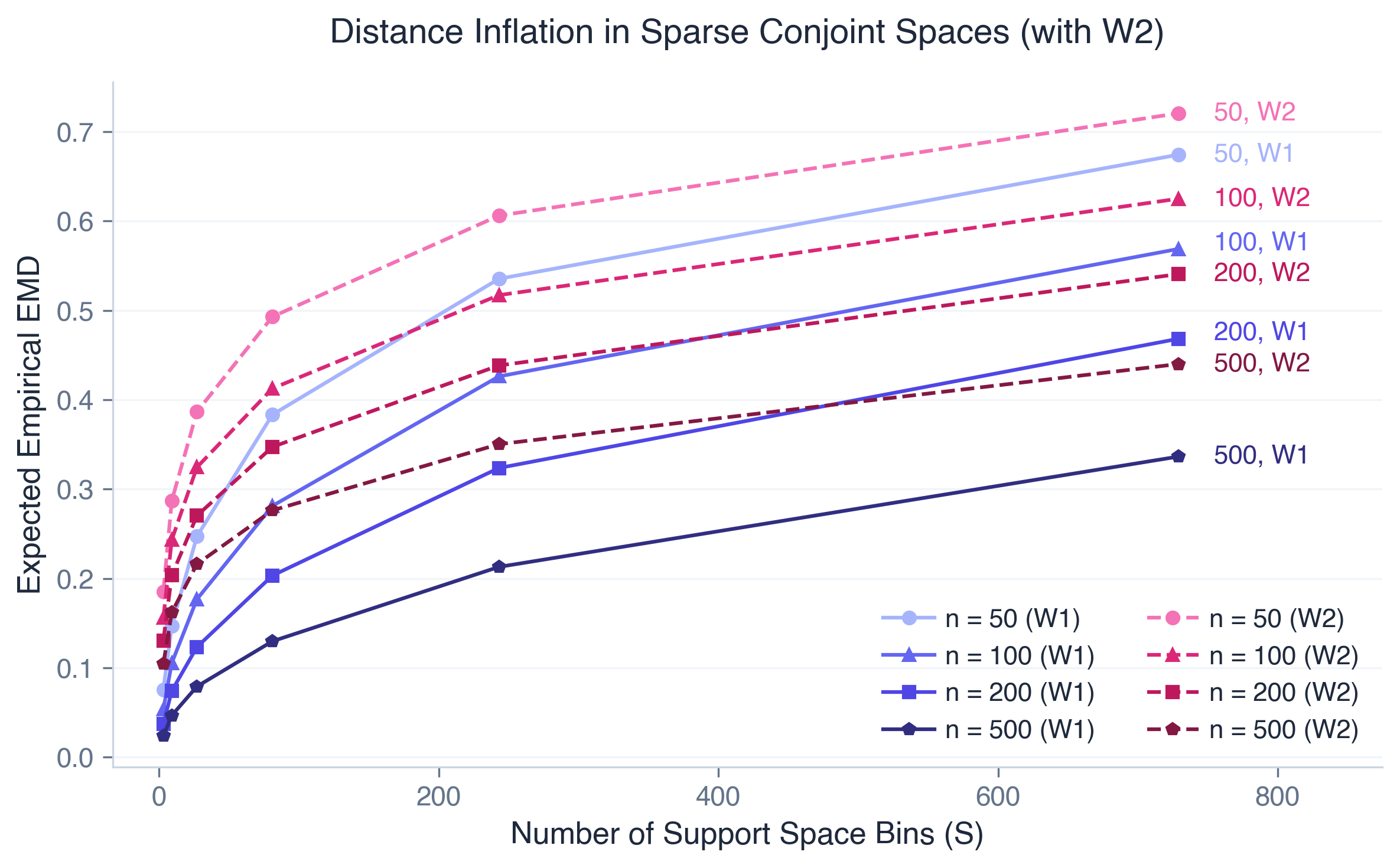}
    \caption{Distance Inflation in Sparse Conjoint Spaces (with $W_2$)}
    \label{fig:3-w2}
\end{figure}

\begin{figure}
    \centering
    \includegraphics[width=\linewidth]{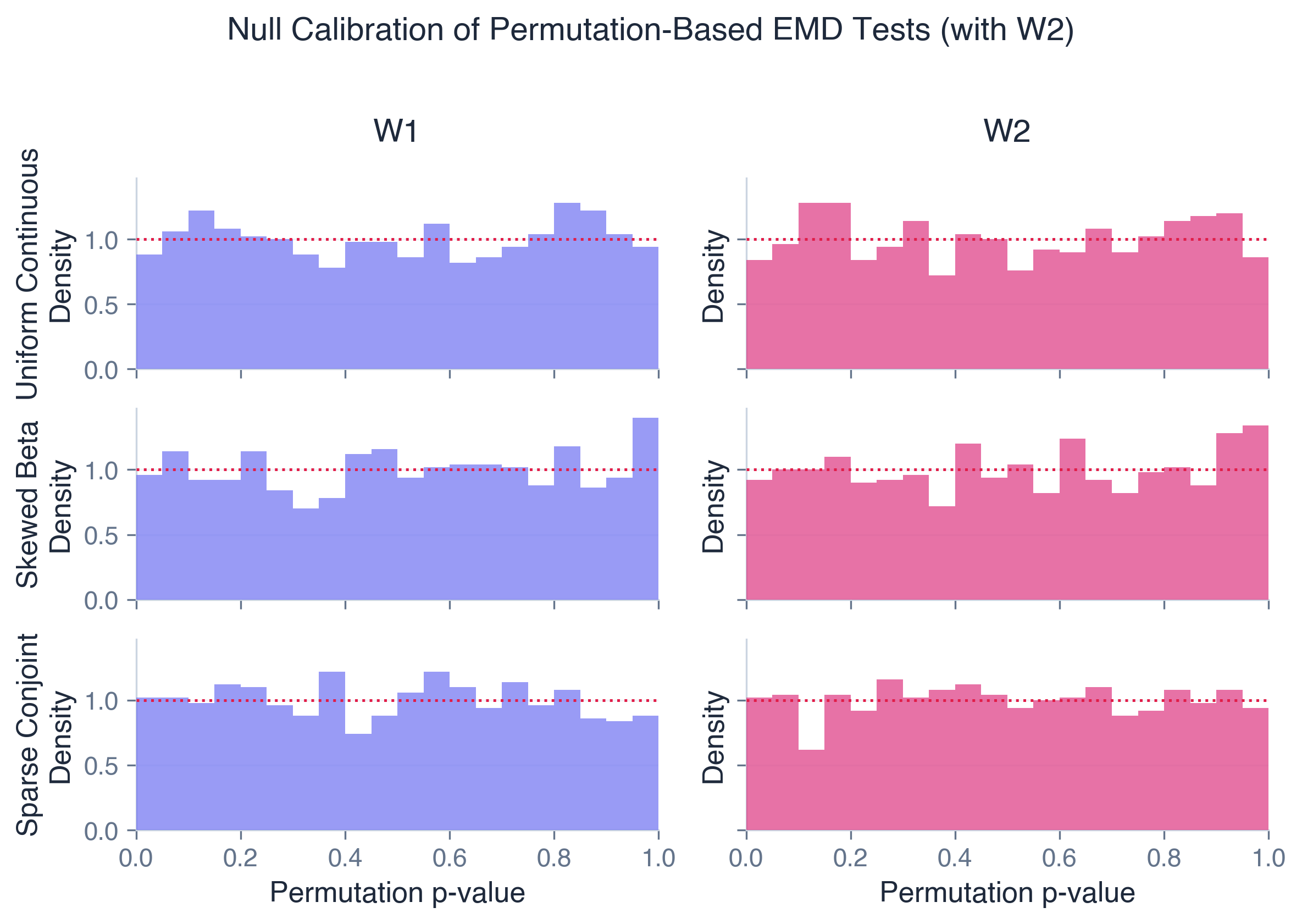}
    \caption{Null Calibration of Permutation-Based EMD Tests (with $W_2$)}
    \label{fig:4-w2}
\end{figure}

The same set of simulations performed in the main body is also done for the case of $W_2$ metric, and the results are reported in Figures~\ref{fig:1-w2}-\ref{fig:4-w2} and Table~\ref{tab:power-with-w2}. The results of $W_1$ are placed adjacent to the results of $W_2$ for easier comparisons. Overall, the upward bias inflation is especially severe for $W_2$ because the squared ground cost places a greater penalty on longer-distance moves. Similar to the $W_1$ case, the procedure here also delivers the expected signal-dependent graded pattern across the settings, which also shows how the proposed approach is robust to the case of $W_2$.

\begin{table}[!htbp]
\centering
\footnotesize
\caption{Size and Power of Permutation-Based $W_1$ and $W_2$ Tests}
\label{tab:power-with-w2}
\begin{tabular}{llcc}
\hline
Setting & Scenario & $W_1$ rejection rate & $W_2$ rejection rate \\
\hline
\multicolumn{4}{l}{\textit{Null scenarios: Type I error}} \\
Continuous & Uniform null & 0.044 & 0.042 \\
Continuous & Skewed Beta null, $\mathrm{Beta}(2,5)$ & 0.048 & 0.046 \\
Sparse conjoint & Uniform null & 0.051 & 0.051 \\
\hline
\multicolumn{4}{l}{\textit{Continuous alternatives: Power}} \\
Continuous & Location shift, $\delta=0.05$ & 0.252 & 0.235 \\
Continuous & Location shift, $\delta=0.075$ & 0.553 & 0.509 \\
Continuous & Location shift, $\delta=0.10$ & 0.829 & 0.796 \\
Continuous & Dispersion, $\mathrm{Beta}(6,6)$ vs. $\mathrm{Beta}(4,4)$ & 0.241 & 0.267 \\
Continuous & Dispersion, $\mathrm{Beta}(7,7)$ vs. $\mathrm{Beta}(4,4)$ & 0.480 & 0.540 \\
Continuous & Dispersion, $\mathrm{Beta}(8,8)$ vs. $\mathrm{Beta}(3,3)$ & 0.990 & 0.993 \\
Continuous & Polarization, $(0.48,0.52)$ vs. $(0.45,0.55)$ & 0.078 & 0.084 \\
Continuous & Polarization, $(0.47,0.53)$ vs. $(0.43,0.57)$ & 0.193 & 0.210 \\
Continuous & Polarization, $(0.44,0.56)$ vs. $(0.38,0.62)$ & 0.784 & 0.792 \\
\hline
\multicolumn{4}{l}{\textit{Sparse conjoint alternatives: Power}} \\
Sparse conjoint & Global shift, $\beta=0.1$ & 0.526 & 0.715 \\
Sparse conjoint & Global shift, $\beta=0.2$ & 1.000 & 1.000 \\
Sparse conjoint & Global shift, $\beta=0.5$ & 1.000 & 1.000 \\
Sparse conjoint & Global shift, $\beta=1.0$ & 1.000 & 1.000 \\
Sparse conjoint & Local one-attribute shift, $\beta=0.1$ & 0.109 & 0.118 \\
Sparse conjoint & Local one-attribute shift, $\beta=0.2$ & 0.354 & 0.472 \\
Sparse conjoint & Local one-attribute shift, $\beta=0.5$ & 1.000 & 1.000 \\
Sparse conjoint & Local one-attribute shift, $\beta=1.0$ & 1.000 & 1.000 \\
\hline
\end{tabular}

\begin{flushleft}
\footnotesize
Note: Under null scenarios, the rejection rate estimates Type I error. Under alternative scenarios, the rejection rate estimates statistical power.
\end{flushleft}
\end{table}

\section{Appendix: Additional Simulation Results with Alternatives}
As an additional calibration benchmark, I compare the empirical Type I error rate of the proposed calibration method with two alternative optimal transport approaches discussed in the main body. Specifically, the comparison evaluates three sets of distinct approaches to the finite-sample inferential problem: directly constructing a permutation-based null reference distribution for the unregularized Wasserstein statistic, a Sinkhorn-divergence statistic calibrated using a pooled non-parametric bootstrap,\footnote{I do not separately report a permutation-calibrated Sinkhorn specification. Under the equality null, the same exchangeability argument underlying the proposed permutation procedure applies to any fixed two-sample statistic, including the Sinkhorn divergence. Therefore, a permutation-calibrated Sinkhorn test cannot act as an independent benchmark of an alternative inferential approach.} as well as a Gaussian $z$-statistic constructed from the centered squared-$W_2$ estimator and dual-potential variance estimator of \textcite{papp_centered_2022}.

The comparisons are conducted for the principal continuous null specification and the principal sparse conjoint specification. Note that the purpose of these comparisons is to benchmark the finite-sample calibration of the proposed permutation procedure against existing alternatives under the main simulation settings. This is different from the objective of the additional robustness specifications, which is to assess the sensitivity of the proposed procedure itself to changes in the data-generating process. They are not intended as a comprehensive comparative simulation study of all competing methods, so the comparisons here are not repeated across every robustness design.

\subsection{Sinkhorn Divergence}
The first alternative statistic is the Sinkhorn divergence, which applies entropic regularization to optimal transport while correcting for the resulting entropic bias. Let $\mathrm{OT}_{\varepsilon}(P,Q)$ denote the full entropy-regularized optimal transport objective between population distributions $P$ and $Q$:
\begin{equation}
\mathrm{OT}_{\varepsilon}(P,Q) = \inf_{\pi \in \Pi(P,Q)} \left\{\int_{\mathcal{X}\times\mathcal{X}} d(x,y)^2 \, d\pi(x,y) + \varepsilon \mathrm{KL} \left(\pi \,\|\, P\otimes Q\right)\right\}
\end{equation}
where $\varepsilon>0$ controls the strength of regularization. The Sinkhorn divergence is then defined as:
\begin{equation}
S_{\varepsilon}(P,Q) = \mathrm{OT}_{\varepsilon}(P,Q) - \frac{1}{2}\mathrm{OT}_{\varepsilon}(P,P) - \frac{1}{2}\mathrm{OT}_{\varepsilon}(Q,Q)
\end{equation}
where the subtraction of the two self-transport terms removes the entropic bias of the regularized transport objective and ensures that the population divergence is zero when $P=Q$. 

In the continuous simulations, I compute the corresponding empirical statistic $S_{\varepsilon}(\hat P_n,\hat Q_n)$ between the two equally weighted empirical point clouds. In the conjoint simulations, I instead compute $S_{\varepsilon}(p,q)$ between the empirical probability vectors over the common $K=729$ profile support. I use the squared-Euclidean ground cost and fix the entropic regularization parameter at $\varepsilon=0.05$.

I calibrate the Sinkhorn statistic using a pooled non-parametric bootstrap. Under $H_0:P=Q$, the two observed groups are pooled to estimate their common distribution. For each Monte Carlo simulation, I draw two bootstrap samples of the original group size independently with replacement from this pooled empirical distribution. In the continuous setting, these produce bootstrap empirical measures $\hat P_n^{*(b)}$ and $\hat Q_n^{*(b)}$; in the conjoint setting, they produce the corresponding bootstrap probability vectors $p^{*(b)}$ and $q^{*(b)}$. The Sinkhorn divergence is then recomputed for each bootstrap replication.

Let $B_{\mathrm{boot}}=500$ denote the number of bootstrap replications, matching the number of resampling iterations used to construct the permutation reference distribution for the proposed procedure.\footnote{This provides a minimum attainable $p$-value increment of approximately 0.002.} Let $S_{\varepsilon}^{\mathrm{obs}}$ denote the observed empirical Sinkhorn divergence. The bootstrap $p$-value is:
\begin{equation}
p_{\mathrm{Sinkhorn}} = \frac{1+\sum_{b=1}^{B_{\mathrm{boot}}} \mathbb{I} \left\{
S_{\varepsilon}^{*(b)} \geq S_{\varepsilon}^{\mathrm{obs}}\right\} }{B_{\mathrm{boot}}+1}
\end{equation}
where $S_{\varepsilon}^{*(b)}$ denotes the Sinkhorn divergence calculated from the $b$th pair of pooled bootstrap samples. The null hypothesis is rejected when $p_{\mathrm{Sinkhorn}}\leq0.05$.\footnote{This pooled $n$-out-of-$n$ bootstrap is used here as a practical finite-sample calibration device. Applying $m$-out-of-$n$ bootstrap could introduce tuning sensitivity. The resulting comparison therefore evaluates the calibration of this particular bootstrap implementation of the Sinkhorn divergence, rather than an intrinsic Type I error property of the Sinkhorn divergence itself. See \textcite{goldfeld_limit_2024} for a detailed discussion of the Sinkhorn null limit distribution.}

\subsection{\textcite{papp_centered_2022} Centered Plug-in Estimator}
The next comparison is based on the centered plug-in estimators introduced by \textcite{papp_centered_2022}. Let $\hat P_n$ and $\hat P_n'$ denote two independent empirical distributions sampled from $P$, and let $\hat Q_n$ denote an empirical distribution sampled from $Q$. I implement the centered statistic:
\begin{equation}
U_n = W_2^2(\hat P_n',\hat Q_n) - W_2^2(\hat P_n',\hat P_n)
\end{equation}
Under the null hypothesis $H_0:P=Q$, the two terms have the same finite-sample expectation, implying $E[U_n]=0$.

Following the uncertainty-quantification construction of \textcite{papp_centered_2022}, I obtain the Kantorovich dual potentials associated with the two optimal transport problems and use their proposed plug-in variance estimator for $U_n$. Under their regularity conditions, \textcite{papp_centered_2022} state that $\sqrt{n}(U_n - E[U_n])$ has a Gaussian limit whose variance is determined by the population Kantorovich potentials. At the equality null $P=Q$, however, this first-order asymptotic variance collapses to the self-transport case, so this first-order asymptotic variance is zero and the $\sqrt{n}$--limit is degenerate. Therefore, the standardized statistic used below should be viewed as a finite-sample Gaussian-$z$ extension of their uncertainty-quantification method, rather than as a hypothesis test whose null distribution is established by \textcite{papp_centered_2022}. The standardized statistic is defined as:
\begin{equation}
Z_n = \frac{U_n}{\widehat{\mathrm{se}}(U_n)}
\end{equation}
The upper-tail Gaussian $p$-value is computed as:
\begin{equation}
p_{\mathrm{PS}} = 1-\Phi(Z_n) 
\end{equation}
where $\Phi(\cdot)$ denotes the standard normal cumulative distribution function. The null hypothesis is rejected when $p_{\mathrm{PS}}\leq0.05$. 

For the sparse conjoint setting, I evaluate this centering construction directly on the empirical probability vectors $p$, $p'$, and $q$ over the common support, using squared Euclidean ground costs over normalized profile coordinates. The plug-in variance estimator and test statistic $Z_n$ are formed analogously from the resulting discrete dual potentials.\footnote{The Papp-Sherlock procedure requires an additional independent empirical sample $\hat P_n'$, whereas the ordinary two-sample procedures only require $\hat P_n$ and $\hat Q_n$. The comparison here therefore concerns Type I error calibration under equal sample size $n$.}

\subsection{Type I Error Analysis}
For each testing procedure $h$, I estimate the empirical Type I error rate as the proportion of the $R=1,000$ null simulations in which the null hypothesis is rejected:
\begin{equation}
\widehat{\alpha}_h = \frac{1}{R} \sum_{s=1}^{R} \mathbb{I} \left\{p_{h,s}\leq0.05\right\}
\end{equation}
where $s$ indexes the Monte Carlo simulation.

Table~\ref{tab:type1_comparison} reports the resulting Type I error rates for the permutation-calibrated $W_1$ and $W_2$ tests, the Gaussian-$z$ implementation of the Papp-Sherlock centered $W_2^2$ procedure, and the pooled-bootstrap calibration of the Sinkhorn divergence. In the continuous case, the proposed permutation procedures are closest to the nominal $0.05$ level, with rejection rates of $0.044$ for $W_1$ and $0.042$ for $W_2$, as reported in the main body earlier. The pooled-bootstrap Sinkhorn implementation rejects in 0.034 of the simulations, indicating modest under-rejection for this bootstrap calibration in the continuous design. The Gaussian-$z$ implementation based on the Papp-Sherlock procedure rejects in only $0.003$ of the simulations. This pattern is consistent with the behavior of its Gaussian uncertainty approximation near the equality null, where the asymptotic variance degenerates and finite-sample variance estimation can become conservative.\footnote{I do not apply an additional bootstrap calibration to the Papp-Sherlock statistic because their proposed uncertainty quantification for the centered estimator $U$ is based on its Gaussian Central Limit Theorem and corresponding variance estimator.} In the conjoint case, the same pooled-bootstrap Sinkhorn implementation rejects in 0.075 of simulations, indicating over-rejection for this calibration scheme. One possible explanation is that, under the equality null, the Sinkhorn divergence has a degenerate first-order limit and $n^{-1}$-order fluctuations. The discrepancy may be amplified by the sparse finite-support structure of the conjoint design. Meanwhile, the Gaussian-$z$ implementation based on the \textcite{papp_centered_2022} procedure remains highly conservative with a rejection rate of $0.003$. The permutation-calibrated tests continue to deliver empirical Type I error rates closest to the nominal $0.05$ level ($0.051$ for both $W_1$ and $W_2$). 

To put it into context, with $R = 1,000$ replications and $\alpha = 0.05$, the Monte Carlo standard error under the null hypothesis is $\sqrt{\frac{0.05 \times 0.95}{1000}} \approx 0.0069$, corresponding to an approximate 95\% simulation CI of $[0.0365, 0.0635]$. The empirical rejection rates of the permutation tests fall entirely within this interval. The rejection rates produced by the Gaussian-$z$ Papp–Sherlock implementation and by the pooled-bootstrap Sinkhorn implementation fall outside it in the implementations considered here. These differences therefore exceed what would ordinarily be attributed to Monte Carlo variation alone. 

While these results do not establish general superiority of one method over another, they indicate that the permutation-calibrated Wasserstein tests maintain Type I error closer to the nominal level under the simulation design considered here.

\begin{table}[!htbp]
\centering
\footnotesize
\caption{Type I Error Rates of Alternative Methods}
\label{tab:type1_comparison}
\begin{tabular}{lc}
\hline
Method & Rejection rate \\
\hline
\multicolumn{2}{l}{\textit{Continuous point cloud}} \\
Gaussian-$z$ implementation based on \textcite{papp_centered_2022} & 0.003 \\
Pooled-bootstrap Sinkhorn implementation & 0.034 \\
\hline
\multicolumn{2}{l}{\textit{Sparse conjoint histogram}} \\
Gaussian-$z$ implementation based on \textcite{papp_centered_2022} & 0.003 \\
Pooled-bootstrap Sinkhorn implementation & 0.075 \\
\hline
\end{tabular}
\end{table}

\end{document}